\documentclass[a4paper,11pt]{article}
\usepackage{amsmath}
\DeclareMathOperator{\plim}{p-lim} 
\usepackage{amsfonts}
\usepackage{tabularx,ragged2e,booktabs,caption}
\usepackage{setspace}
\usepackage{amssymb}
\RequirePackage{natbib}
\usepackage{graphicx}
\usepackage{graphics}
\usepackage{enumitem}
\usepackage{array}
\newcolumntype{L}[1]{>{\raggedright\arraybackslash}p{#1}}
\newcolumntype{C}[1]{>{\centering\arraybackslash}p{#1}}
\newcolumntype{R}[1]{>{\raggedleft\arraybackslash}p{#1}}
\usepackage{geometry}
\usepackage{booktabs}
\usepackage{tabularx, array}
\usepackage{rotating}
\usepackage{dcolumn}

\title{\textbf{Modelling and Forecasting the Realized Range Conditional Quantiles}}
\author{Giovanni Bonaccolto \footnote{Dipartimento di Scienze Statistiche, Via Cesare Battisti 241, 35121 Padova, Italy. Email: bonaccolto@stat.unipd.it.} \\
\and
Massimiliano Caporin \footnote{Dipartimento di Scienze Economiche e Aziendali \textit{Marco Fanno}, Via del Santo 22, 35123 Padova, Italy. 
Email: massimiliano.caporin@unipd.it.}} 
\date{}
\begin{document}
\maketitle

\singlespacing
\begin{abstract}
Several studies have focused on the Realized Range Volatility, an estimator of the quadratic variation of financial prices, taking into account the impact of microstructure noise and jumps. However, none has considered direct modeling and forecasting of the Realized Range conditional quantiles. This study carries out a quantile regression analysis to fill this gap. The proposed model takes into account as quantile predictors both the lagged values of the estimated volatility and some key macroeconomic and financial variables, which provide important information about the overall market trend and risk. In this way, and without distributional assumptions on the realized range innovations, it is possible to assess the entire conditional distribution of the estimated volatility. This issue is a critical one for financial decision-makers in terms of pricing, asset allocation, and risk management. The quantile regression approach allows how the links among the involved variables change across the quantiles levels to be analyzed. In addition, a rolling analysis is performed in order to determine how the relationships that characterize the proposed model evolve over time. The analysis is applied to sixteen stocks issued by companies that operate in differing economic sectors of the U.S. market, and the forecast accuracy is validated by means of suitable tests. The results show evidence of the selected variables’ relevant impacts and, particularly during periods of market stress, highlights heterogeneous effects across quantiles.

\textbf{Keywords}: Forecast assessment, Principal component analysis, Quantile regression, Realized Range Volatility, Rolling analysis, Volatility quantiles forecasting.
\end{abstract}


\section{Introduction}
\singlespacing

Recent events, such as the subprime crisis, which originated in the United States and was marked by Lehman Brothers’ default in September 2008, and the sovereign debt crisis, which hit the Eurozone in 2009, have highlighted the fundamental importance of risk measurement, monitoring, and forecasting. The volatility of asset returns, a commonly used measure of risk, is a key variable in several areas of finance and investment, such as risk management, asset allocation, pricing, and trading strategies. Therefore, estimating and forecasting the volatility point values and distribution play a critical role. The use of predicted volatility levels is central, for instance, in the pricing of equity derivatives, in the development of equity derivative trading strategies, and in risk measurement when risk is associated with volatility, while the volatility distribution is of interest for trading/pricing volatility derivatives, for designing volatility hedges for generic portfolios, and for accounting for the uncertainty on volatility point forecasts.

Many financial applications use constant volatility models \citep{BlSc73}, although empirical evidence suggests that variance changes over time. Several approaches have been developed with the purpose of achieving more accurate estimates, such as the class of ARCH \citep{En82} and GARCH \citep{Bo86} models and the stochastic volatility models \citep{Metu90, Ta94, HaRuSh94, JaPoRo94}. Nevertheless, financial data are affected by several features, and the so-called stylized facts \citep{Co01} and standard GARCH and stochastic volatility models do not capture all of them \citep{Co09}. 

We might also estimate volatility with non-parametric methods, such as by means of realized measures, which have been shown to perform better than traditional GARCH and stochastic volatility models when forecasting conditional second-order moments \citep{AnBoDiLa03}. This approach has attracted considerable interest because of the availability of high-frequency financial data. In fact, as opposed to models that treat volatility as a latent (non-observable) element, realized measures use additional information from the intraday returns. Moreover, by resorting to observable data and extracting volatility by a non-parametric, model-free approach, realized measures avoid the risk of model misspecification. We believe these methods provide more flexibility than standard GARCH-type models do, so we focus on realized measures here.

The realized variance \citep{AnBoDiLa03, BaNiSh02}, which is based on continuous time price theory and computed starting from the sum of squared intraday returns, is the natural estimator of the ex-post integrated volatility. In particular, the realized variance is an unbiased estimator of the quadratic variation and converges to it as the sampling frequency rises to infinity \citep{Ja94, JaPr98, BaNiSh02}. \cite{MaVD07} and \cite{ChPo07} proposed the realized range-based variance, an estimator that increases efficiency by replacing every squared return of the realized variance with a normalized squared range using the maximum and minimum prices observed in each of the subintervals into which the trading day is split.

The presence of microstructure noise becomes a problem that must be also considered in the context of high-frequency financial data. The microstructure noise that arises from peculiar phenomena like non-continuous trading, infrequent trades, and bid-ask bounce \citep{Ha06, OH98, Ro84} affects the properties of the estimators. The literature offers several corrections. One consists of reducing the sampling frequency; for instance \cite{BaRu06} suggested the rule of thumb of using five-minutes returns, which optimizes the bias-variance trade-off. Another solution refers to the subsampling of two time-scales realized variance \citep{ZhMyAS05}, while a third refers to the kernel estimation introduced by \cite{HaLu06}. Besides the microstructure noise, the volatility of financial returns is affected by large and rapid increments. In particular, \cite{ErJoPo03} showed that jumps in returns could imply large movements whose impacts are transient. Jumps in volatility, on the other hand, are rapid but persistent factors that drive volatility. 

Among the various realized measures, the range-based measure is the most efficient, although microstructure noise and jumps must be taken into account. In order to ensure accurate results, \cite{ChPoVe09} proposed the realized range-based bias corrected bipower variation based on adjustments made by the variance of the noise and a constant computed through simulation methods, which \cite{ChPoVe09} demonstrated is a consistent estimator of the integrated variance in the presence of noise and jumps. Given its efficiency, we chose the realized range-based bias corrected bipower variation as the reference estimator for daily integrated volatility.

Many studies that focus on volatility forecasting have identified through several approaches key macroeconomic and financial variables as important drivers of volatility, highlighting their power in improving forecast performances. For instance, \cite{ChScSc12} predicted the asset return volatility by means of macroeconomic and financial variables in a Bayesian Model Averaging framework. They considered several asset classes, such as equities, foreign exchange, bonds, and commodities, over long time spans and found that economic variables provide information about future volatility from both an in-sample and an out-of-sample perspective. \cite{Pa12} tested the power of financial and economic variables to forecast the volatility at monthly and quarterly horizons and rarely found a statistical difference between the performance of macroeconomic fundamentals and univariate benchmarks. \cite{FeMeSc09} used parametric and semi-parametric Heterogeneous Auto Regressive (HAR) processes to model and forecast the VIX index and found significant results using financial and macroeconomic variables as additional regressors. \cite{CaVe11} used an HAR model with asymmetric effects with respect to volatility and return, and GARCH and GJR-GARCH specifications for the variance equation. \cite{CaRoMa11} studied the relationship between the first principal component of the volatility jumps, estimated using thirty-six stocks and a set of macroeconomic and financial variables, such as VIX, S\&P 500 volume, CDS, and Federal Fund rates, and found that CDS captures a large part of the moves of the expected jumps. Given the findings of these studies, we, too, use macroeconomic and financial variables in our model.  

Our purpose is to generalize the previous contributions further. We model and forecast the conditional quantiles of the realized range-based bias corrected bipower variation by means of the quantile regression method introduced by \cite{KoBa78}. Other authors have applied quantile regression in a financial framework. For instance, \cite{EnMa04} proposed the CAViaR model to estimate the conditional Value-at-Risk, an important measure of risk that financial institutions and their regulators employ. For their part, \cite{WhKiMa08} generalized the CAViaR to the Multi-Quantile CAViaR (MQ-CAViaR) model, studying the conditional skewness and kurtosis of S\&P 500 daily returns. \cite{WhKiMa10} extended the MQ-CAViaR model in the multivariate context to measure the systemic risk, a critical issue highlighted by the recent financial crises, taking into account the relationships among 230 financial institutions from around the world. Finally, \cite{CaPeRaRi14} adopted quantile regressions as a tool with which to detect financial contagion across bond spreads in Europe. 

Focusing on realized volatility measures, we believe that the quantile regression approach can provide useful new evidence by allowing the entire conditional distribution of the realized volatility measure to be estimated instead of restricting the attention to the conditional mean. This approach could have a relevant advantage when the impact of covariates is changing, depending on market conditions (say, on low or high volatility states), or when the purpose is to recover density forecasts without making a distributional assumption. 

\cite{ZiBa13} used a similar approach to estimate the conditional quantiles of volatility, but our work differs in some important ways. First of all, \cite{ZiBa13} estimated volatility by means of a realized measure that takes into account only the effects of jumps in the price process. However, we use the realized range-based bias corrected bipower variation, which considers the impact of microstructure noise as well as that of jumps. What’s more, our approach is based on squared ranges, rather than on squared returns, so it has the advantage of using additional information. To the best of our knowledge, quantile regression methods for the analysis of realized range volatility measures have never been used in the econometric and empirical financial literature, which provides a strong motivation for our study. Secondly, \cite{ZiBa13} used a heterogeneous autoregressive quantile model, whereas we also made use of conditioning exogenous variables. Third, \cite{ZiBa13} performed quantile forecasts, focusing on a few quantiles, but  we go farther by considering the entire conditional distribution of volatility. We stress that using interpolation to recover the density forecasts from a collection of quantile forecasts does not require a distributional assumption, as HAR-type models do, and to the best of our knowledge, such an approach has never been used in financial applications. Finally, we also evaluated the structural changes in our model over time by means of a rolling analysis.

Our work takes an empiric point of view and focuses on the high-frequency data of sixteen stocks issued by large-cap companies that operate in differing economic sectors. All companies are quoted on the U.S. market. In a first step, we analyze the first principal component of the estimated volatility as a summary of the sixteen series (a kind of market factor). 

We provide a first contribution to the literature by showing that some macro-finance-related variables have a significant impact on volatility quantiles; that their impact changes across quantiles, becoming irrelevant in some cases; and that the significance and strength of the relationship changes over time. This last finding is particularly evident when we focus on turbulent market phases, as in these periods we note an increase in the impact of some variables, such as the VIX. The last finding is particularly evident for high-volatility quantiles. The heterogeneity we observe across quantiles can also be interpreted as evidence against the location-shift hypothesis. 

Our second contribution comes from the single asset analyses. We develop a specific model for each asset in order to determine how the features of the sixteen companies affect the relationships among the variables involved. We find some heterogeneity in the assets’ reactions to the macro-finance variables, which holds across both time and volatility quantiles. However, we find an overall confirmation of the findings associated with the first principal component.

A third contribution of our analyses stems from a forecasting exercise. We compare the quantile-based density forecasts, which we recover by interpolating a set of volatility quantiles, to those of a benchmark model adapted to the realized range volatility mean and variance. The reference model combines a HAR  structure on the realized volatility mean, plus a GJR-GARCH \citep{GlJaRu93} for the mean innovation variances. The HAR structure, inspired by the work of \cite{Co09}, captures the persistence of realized measures and is consistent with the presence of heterogeneous agents in the market, while the GJR-GARCH is coherent with \cite{CoMiPiPi08}’s volatility of volatility hypothesis. We compare the benchmark model density forecasts and the quantile regression-based forecasts by means of the tests proposed by \cite{Be01} and \cite{AmGi07}. Moreover, by using a quantile-based loss function, we also consider the \cite{DiMa02} test. We stress that the \cite{Be01} test allows for an absolute evaluation of the density forecasts provided by one model, while the \cite{AmGi07} approach compares two competing models. The results confirm that our approach performs better, thus providing support for the use of quantile regression methods in all areas where volatility quantiles might have a role. Among the possible applications, we mention volatility trading and volatility hedging. 

The remainder of the paper is structured as follows. Section 2 includes the description of the data. Section 3 presents the model we propose to forecast the range bipower variation conditional quantiles. Section 4 is devoted to the density forecast and predictive accuracy and provides the details about the tests we use. The results are analyzed in Section 5, and Section 6 provides a set of concluding remarks.

\section{Data description}

The database we use includes stock prices recorded with a frequency of one minute, from 9:30 a.m. to 4:00 p.m. of every trading day between January 2, 2003, and June 28, 2013, inclusive. The equities analyzed are those of large companies that operate in various economic sectors of the U.S. market: AT\&T Inc. (\textit{ATT}), Bank of America (\textit{BAC}), Boeing (\textit{BOI}), Caterpillar, Inc. (\textit{CAT}), Citigroup, Inc. (\textit{CTG}), FedEx Corporation (\textit{FDX}), Honeywell International, Inc. (\textit{HON}), Hewlett-Packard Company (\textit{HPQ}), International Business Machines Corp. (\textit{IBM}), JPMorgan Chase \& Co. (\textit{JPM}), Mondelez International, Inc. (\textit{MDZ}), Pepsico, Inc. (\textit{PEP}), The Procter \& Gamble Company (\textit{PRG}), Time Warner, Inc. (\textit{TWX}), Texas Instruments, Inc. (\textit{TXN}), and Wells Fargo \& Company (\textit{WFC}). 

The dataset is drawn from TickData. The prices are adjusted for extraordinary operations and filtered for errors, anomalies, and outliers that arise from traders' activities.\footnote{For additional details see the company website: \textit{http://www.tickdata.com}.} Table \ref{tab:descr} shows the most relevant descriptive statistics of the daily returns from the companies under consideration. The values confirm some of the stylized facts that characterize the financial data \citep{Co01}. The distributions of the asset returns have heavy tails, as expected: \textit{TXN} has the smallest value of kurtosis, while the largest is recorded for \textit{CTG}. The skewness index is less than zero for twelve of the companies, which is consistent with the so-called gain/loss asymmetry phenomenon. The mean and the median are close to zero for all assets considered. The standard deviation ranges from 0.011 to 0.0368, while the interquartile range goes from 1.06\% to 2.27\%.

\begin{table}[htb]
\begin{center}
\centering
\small
\captionof{table}{Descriptive statistics of the daily returns.}
\begin{tabular}{ccccccc}
\hline
Stock & Mean & St. Dev. & Median & IQR & Skewness & Kurtosis \\
\hline
ATT & 0.0001 &    0.0151 &   0.0004  &  0.0145   & 0.3528  & 10.8522\\
BAC & -0.0004 & 0.0348 &  0.0000  & 0.0188 &   -0.3792 & 25.4668\\
BOI & 0.0004 & 0.0183 & 0.0005 & 0.0196 & 0.1323 & 7.3454 \\
CAT & 0.0005 & 0.0211 & 0.0006 & 0.0213 & -0.1710 & 8.0297\\
CTG & -0.0008 & 0.0368 & 0.0000 & 0.0203 & -0.8930 & 41.8702\\
FDX & 0.0002  &  0.0190 &        0.0000   & 0.0194   & -0.2412  &  7.6775 \\
HON & 0.0004  &  0.0178  &  0.0003 &   0.0183 &   -0.0854 &    6.6231\\
HPQ & 0.0001 &  0.0213 &    0.0004 &   0.0207 &   -0.6265 &   14.8602\\
IBM & 0.0003 & 0.0138 &    0.0002  &  0.0138   & -0.1195  &  8.7373\\
JPM & 0.0003 &   0.0265     &    0.0000   & 0.0197    & 0.2925  & 17.7241\\
MDZ & 0.0000 &    0.0130  &  0.0001   &  0.0127 &  -0.6790 &  13.4781\\
PEP & 0.0002 &   0.0112    &     0.0000 &   0.0110 &   -0.4096 &   16.9905\\
PRG & 0.0002 &    0.0110   &  0.0002   & 0.0106   & -0.1384  & 10.4464\\
TWX & 0.0003   & 0.0189     &    0.0000  &  0.0183  & -0.1137  & 11.6788\\
TXN & 0.0003  &  0.0207  &  0.0003  &  0.0227   & -0.0650  &  6.3780\\
WFC & 0.0002  &  0.0284     &    0.0000  &  0.0168    & 0.7318 &  26.4605\\
\hline
\end{tabular}\par
\label{tab:descr} 
\end{center}
\footnotesize{The table reports for each stock (the ticker is given in the first column) a set of descriptive statistics for the daily returns. IQR stands for Inter-Quartile Range.}
\end{table}

While the daily returns are computed from the close prices of each trading day, the volatilities of the sixteen assets for each day are estimated through the realized range-based bias corrected bipower variation, $RRV^{n,m}_{BVBC}$, introduced by \cite{ChPoVe09}.\footnote{See Appendix A for the details on the methodology we adopted for the integrated volatility estimation.} We checked that the series of the estimated volatilities are affected by the 2008-2009 financial crisis and that this phenomenon is particularly noticeable in the case of the financial companies. 

A principal component analysis is carried out on the range-based bias corrected bipower variations of the sixteen assets. The evolution of the assets’ volatilities have a strong common behavior that we might interpret as \textit{market} or \textit{systematic} behavior.
In particular, the first principal component ($FPC$) explains 77\% of the overall variation, therefore analysis of the first principal component could produce useful results.
We observed that  $FPC$ is strongly affected by the 2008-2009 financial crisis since it shows high values in that period. Second, these peaks are extreme values in the right tail of the $FPC$ distribution, suggesting the wisdom of using quantile regression rather than regression on the mean because regression on the mean is particularly sensitive to extreme values. The $FPC$ autocorrelations keep high values for all twenty lags, indicating some persistence, a common finding on realized range/variance sequences.

In addition to the data described so far, our analyses take into account some key macroeconomic and financial variables that convey important information about the overall market trend and risk and that will be considered exogenous variables that affect the conditional quantiles. Other studies \citep{CaVe11, CaRoMa11} have used several indicators to analyze the realized variance and range series, among which we select just two (since the others, such as the logarithmic returns of the U.S. dollar-Euro exchange rate and of oil, were not significant in the present analysis): the daily return of the S\&P 500 index ($sp500$), which reflects the trend of the U.S. stock market, and the logarithm of the VIX index ($vix$), a measure of the implied volatility of S\&P 500 index options. We expected that negative returns of the S\&P 500 would have a positive impact on the market volatility, the well-known leverage effect, while high levels of $vix$ reflect pessimism among the economic agents, so a positive relationship between $vix$ and the volatility level is expected. The observations associated with $sp500$ and $vix$ are also recorded at a daily frequency and are recovered from Datastream. Table \ref{descr_regr} lists some of their most important descriptive statistics. In particular, the distribution of $sp500$ is centered at zero, whereas the mean and the median of $vix$ are equal to 2.94 and 2.89, respectively. $vix$ has greater volatility than $sp500$ does, as the standard deviations and the inter-quartile ranges indicate. Moreover, $sp500$ has a leptokurtic distribution with negative asymmetry, while $vix$ has a lower kurtosis (3.67) and a positive skewness (0.85). 

\begin{table}[htb]
\begin{center}
\centering
\small
\captionof{table}{Descriptive statistics of \textit{sp500} and \textit{vix}.}
\begin{tabular}{ccccccc}
\hline
Variable & Mean & St. Dev. & Median & IQR & Skewness & Kurtosis \\
\hline
sp500 & 0.0003 & 0.0129 & 0.0008 & 0.0108 & -0.0684 & 13.6343 \\
vix & 2.9437 & 0.3845 & 2.8893 & 0.5030 &   0.8512 & 3.6669 \\
\hline
\end{tabular}\par
\label{descr_regr} 
\end{center}
\footnotesize{The table reports some descriptive statistics computed for the return of the S\&P 500 index ($sp500$) and the logarithmic of the VIX index ($vix$), recorded with a daily frequency; IQR stands for Inter-Quartile Range.}
\end{table}

Many macroeconomic and financial time series are not stationary and are often characterized by unit-root non-stationarity \citep{NePl82}. Building on this evidence, we must determine whether the data-generating process of $vix$ is affected by unit root.\footnote{We do not report the tests on $sp500$, as they provided no useful results: the index level is non-stationary, while the index return is stationary.} To this purpose, we consider two standard tests: the augmented Dickey-Fuller (ADF) test \citep{DiFu81} and the Phillips-Perron (PP) test \citep{PhPe88}. The ADF test rejects the null hypothesis at the 10\% level, and the test statistic equals -3.18 (with a p-value of 0.09). On the other hand, the PP test rejects the null hypothesis at the 5\% confidence level, and the test statistic is -24.48 (with a p-value of 0.03). Therefore, we have a moderate amount of evidence against the presence of a unit root for the $vix$ series. The rejection of the null is much clearer with the PP test.


\section{The model}
We now introduce the model we propose to study the conditional quantiles of our realized range volatility measure. We focus, at least in the initial step, on the first principal component of the sixteen assets’ realized range volatilities. We denote by $fpc_t$ the first principal component’s observed value at day $t$. 
In addition to $vix$ and $sp500$, mentioned in Section 2, we use other quantities to describe the conditional quantiles. Those are derived from the available data, so they are not included in the previous section. First, we follow \cite{Co09} in introducing among the explanatory variables those commonly adopted in (\textit{HAR}) models of Realized Volatility. The first one is the lagged value of $fpc_t$, that is, $fpc_{t-1}$. This variable is usually accompanied by other quantities that are built from local averages of past elements:

\begin{equation}\label{fpc5}
\overline{fpc_m}_t=\frac{1}{m}\sum_{i=0}^{m-1}fpc_{t-i}.
\end{equation}

\cite{Co09}’s model used $m=5$ and $m=21$, representing the weekly and monthly horizons. These components allow the heterogeneous nature of the information arrivals \citep{Co09} in the market to be considered. In fact, many operators have differing time horizons. For instance, intraday speculators have a short horizon, while insurance companies trade much less frequently. Therefore, agents whose time horizons differ, perceive, react to, and cause different types of volatility components. In our study we use only $m=5$ with the first lag. In fact, the longer horizon component, with $m=21$, was not significant. We might interpret $\overline{fpc_5}_{t-1}$ as reflecting the medium-term investors who typically rebalance their positions at a weekly frequency. We found that $fpc_{t-1}$ and $\overline{fpc_5}_{t-1}$ are positively correlated with $fpc_t$, suggesting a positive impact at least in the mean.

The last explanatory variable is $jump_{t-1}$, a quantity that takes into account the impact of lagged jumps in the price process. At the single asset level, the jump detection is carried out through the test proposed by \cite{RePEc:zbw:sfb475:200637} based on the comparison between the realized range-based variance, $RRV^{n,m}$, and the range-based bipower variation,  $RBV^{n,m}$.\footnote{See Appendix A for the details on the test and the two estimators.} $RRV^{n,m}$ and $RBV^{n,m}$ are jump non-robust and jump robust estimators of the integrated variance, respectively. We observed that, in periods of financial turmoil, the series of the differences between $RRV^{n,m}$ and $RBV^{n,m}$ have evident peaks, but elsewhere the differences are irrelevant.

Table \ref{jump_testing_ris} lists some important results from computing the test statistic proposed by \cite{RePEc:zbw:sfb475:200637}, $Z_{TP,t}$, for each asset and trading day. For each asset, the percentage of rejections is computed by dividing the number of days in which the null hypothesis of no jumps is rejected by the total number of considered days, given $\alpha=0.01$. That percentage ranges from 44.89\% ($CTG$) to 91.97\% ($MDZ$). The mean of $Z_{TP,t}$ ranges from 2.20 ($CTG$) to 7.85 ($MDZ$), while the minimum and maximum values of the median are 2.12 ($CTG$) and 5.78 ($MDZ$), respectively. Finally, the standard deviation goes from 1.95 ($HPQ$) to 5.68 ($MDZ$). The results given in Table \ref{jump_testing_ris} highlight the importance of jumps in the price process of the sixteen stocks considered.

\begin{table}[htb]
\begin{center}
\centering
\small
\captionof{table}{Jump testing results ($\alpha=1\%)$.} 
\begin{tabular}{ccccc}
\hline
Stock & Mean of $Z_{TP,t}$ & Median of $Z_{TP,t}$  & St. Dev. of $Z_{TP,t}$ & Percentage of rejections  \\
\hline
ATT & 3.9481 & 3.5437 & 2.4303 & 80.87\% \\
BAC & 3.0540 & 2.4240 & 2.6882 & 51.67\% \\
BOI & 4.0296 & 3.5337 & 2.4356 & 73.75\% \\
CAT & 3.7747 & 2.9523 & 2.8838 & 61.63\% \\
CTG & 2.1973 & 2.1218 & 2.5268 & 44.89\% \\
FDX & 5.6218 & 4.5568 & 3.6689 & 85.72\% \\
HON & 4.5356 & 4.1011 & 2.5571 & 83.41\% \\
HPQ & 3.5559 & 3.3228 & 1.9524 & 75.57\% \\
IBM & 3.3777 & 3.0468 & 2.0277 & 69.43\% \\
JPM & 2.8338 & 2.4880 & 2.0705 & 53.86\% \\
MDZ & 7.8518 & 5.7816 & 5.6762 & 91.97\% \\
PEP & 4.6374 & 4.1476 & 2.6429 & 84.28\% \\
PRG & 4.3166 & 3.9646 & 2.3446 & 82.05\% \\
TWX & 3.7861 & 3.3687 & 2.4973 & 75.04\% \\
TXN & 3.5967 & 3.1302 & 2.4412 & 70.87\% \\
WFC & 4.0498 & 3.4154 & 3.1662 & 67.61\% \\
\hline
\end{tabular}\par
\label{jump_testing_ris} 
\end{center}
\footnotesize{The table reports for each stock (the ticker is given in the first column) some results obtained from the ratio-statistic $Z_{TP}$, computed for each day $t$. The percentage of rejections is computed dividing the number of days in which the null hypothesis of no jumps is rejected by the total number of considered days.}
\end{table}

One of the most relevant limits of $RRV^{n,m}$ is that it does not take into account the presence of jumps in the price process. This gap could be remedied by estimating the integrated variance through $RBV^{n,m}$, but we used $RRV^{n,m}_{BVBC}$ in the present work because it is a consistent estimator of the integrated variance in the presence of both microstructure noise and jumps. Therefore, we used the difference $RRV^{n,m}-RRV^{n,m}_{BVBC}$ as a measure of the jump component, denoted for the $i$-th asset at day $t$ as $jump_{i,t}$. We focus first on the first principal component of the realized range measures, so for reasons of consistency, the jump explanatory variable, denoted by $jump_t$, is also the first principal component of the sixteen asset-specific jumps.

To summarize, the variables involved in the model of interest are: $fpc_t$, $fpc_{t-1}$, $\overline{fpc_5}_{t-1}$, $vix_{t-1}$, $sp500_{t-1}$, and $jump_{t-1}$, and their linear correlation coefficients are given in Table \ref{corr_matr}. $fpc_{t-1}$ has the highest correlation with $fpc_t$, and the signs of the correlation coefficients are consistent with expectations, since $fpc_{t-1}$, $\overline{fpc_5}_{t-1}$, $vix_{t-1}$, and $jump_{t-1}$ are positively correlated with $fpc_t$, while $sp500_{t-1}$ has a negative correlation coefficient. 

\begin{table}[htb]
\begin{center}
\centering
\small
\captionof{table}{Linear correlation coefficients.}
\begin{tabular}{ccccccc}
\hline
 & $fpc_t$ & $fpc_{t-1}$ & $\overline{fpc_5}_{t-1}$ & $vix_{t-1}$ & $sp500_{t-1}$ & $jump_{t-1}$ \\
\hline
$fpc_t$ & 1.00 & 0.75 & 0.73 & 0.54 & -0.14 & 0.50 \\ 
$fpc_{t-1}$ & 0.75 & 1.00 & 0.83 & 0.54 & -0.05 & 0.55 \\
$\overline{fpc_5}_{t-1}$ & 0.73 & 0.83 & 1.00 & 0.62 & 0.01 & 0.59 \\
$vix_{t-1}$ & 0.54 & 0.54 & 0.62 & 1.00 & -0.11 & 0.36 \\
$sp500_{t-1}$ & -0.14 & -0.05 & 0.01 & -0.11 & 1.00 & -0.03 \\
$jump_{t-1}$ & 0.50 & 0.55 & 0.59 & 0.36 & -0.03 & 1.00 \\
\hline
\end{tabular}\par 
\label{corr_matr} 
\end{center}
\footnotesize{The table reports the full-sample linear correlation coefficients across the variables entering in model (\ref{mainmodel}).}
\end{table}

All of these variables enter a linear specification for conditional quantiles of the first principal component of realized volatility measures, so the model we consider is

\begin{eqnarray}\label{mainmodel}
Q_{fpc_t}(\tau|\mathbf{x}_{t-1}) & = & \mathbf{x}_{t-1}'\pmb{\beta}(\tau)=\beta_0(\tau)+\beta_1(\tau)fpc_{t-1}+\beta_2(\tau)\overline{fpc_5}_{t-1}+\beta_3(\tau)vix_{t-1} \nonumber \\ & + & \beta_4(\tau)sp500_{t-1}+\beta_5(\tau)jump_{t-1},
\end{eqnarray}

where $Q_{fpc_t}(\tau|\mathbf{x}_{t-1})$ denotes the $\tau$th quantile of $fpc_t$, conditional to the information included in $\mathbf{x}_{t-1}$. Although this approach is not novel, as conditional quantiles have already been used in a risk-management framework (e.g., \cite{EnMa04}, \cite{WhKiMa08} and \cite{WhKiMa10}), we stress that, to the best of our knowledge, realized measures based on ranges have never been used. Therefore, even a simple estimation of model (\ref{mainmodel}) would provide useful results in terms of revealing the impact of covariates and the stability of the various coefficients across quantiles. Quantile regressions could also be used to forecast the conditional quantiles of the realized range volatility sequence, as we discuss below.\footnote{See Appendix B for details on quantile regression estimation and testing.}

Certain events, such as the subprime crisis that was born in the U.S. and that was marked by Lehman Brothers’ default in September 2008 and the Eurozone sovereign debt crisis in 2010-2011, had considerable effects on the mechanisms that govern the international financial system. These extreme events could have affected the relationship between control variables and conditional quantiles, so it is necessary to determine whether the relationships that characterize model (\ref{mainmodel}) change over time before, during, or after these period of turmoil. For this purpose, we performed a rolling analysis with a step of one day and a window size of 500 observations. Thus, it is possible to determine how the coefficients’ values evolve over time and over $\tau$.
We also built model (\ref{mainmodel}) to predict the conditional quantiles of the first principal component computed on the realized range-based bias corrected bipower variations of the sixteen assets. The forecasts, produced for a single step ahead, provide relevant details for the covariates’ prediction abilities. 

As the underlying companies have differing features and operate in differing economic sectors, it is also useful to build a model for each asset. These asset-specific models have the same structure of that one given in equation (\ref{mainmodel}), but the dependent variable is the conditional $RRV_{BVBC}^{n,m}$ quantile of the one asset, and $fpc_{t-1}$, $\overline{fpc_5}_{t-1}$, and $jump_{t-1}$ are replaced with the analogous quantities computed for each asset. Therefore, the model built for the $i$-th asset, for $i=1,...,16$, is

\begin{eqnarray}\label{single_asset_model}
Q_{rrv_{i,t}}(\tau|\mathbf{x}_{i,t-1}) & = & \beta_{0,i}(\tau)+\beta_{1,i}(\tau)rrv_{i,t-1}+\beta_{2,i}(\tau)\overline{rrv_5}_{i,t-1}+\beta_{3,i}(\tau)vix_{t-1} \nonumber \\ & + & \beta_{4,i}(\tau)sp500_{t-1}+\beta_{5,i}(\tau)jump_{i,t-1},
\end{eqnarray}

where $rrv_{i,t}$ is the observed $RRV_{BVBC}^{n,m}$ related to the $i$-th company at day $t$, $\overline{rrv_5}_{i,t}$ is the mean of the $rrv_{i,t}$ values recorded in the last 5 days, and $jump_{i,t-1}$ is the difference between $RRV^{n,m}$ and $RRV_{BVBC}^{n,m}$, computed for the $i$-th stock at day $t-1$.

The models we propose allow the conditional quantiles of a realized volatility measure to be estimated. We don't know the true, unobserved volatility of the assets returns (they are estimated) so measurement errors might play a role. However, we restrict our attention to the forecast of the realized measure at a given sampling frequency, not a forecast of the future returns volatility. As a consequence, as \cite{ZiBa13} discussed, the impact of measurement errors has limited importance.

\section{Density forecast and predictive accuracy}

Equation (\ref{mainmodel}) provides the expression for the point forecast of a volatility quantile, which can be used to assess the forecasting performances of the model for specific points on the volatility density. However, while the entire volatility density could be of interest if one is dealing with volatility trading or volatility hedging applications, it  is not readily available from a quantile regression approach. Still, it can be recovered by approximation by interpolating a large set of volatility quantile forecasts. Notably, such a method does not require a distributional assumption for the innovations of the conditional quantiles. Therefore, the density forecast is recovered using a semi-parametric approach. In this case, the analysis takes a density-forecasting perspective, where assessment of a proposed approach’s predictive power (as compared to a benchmark model) and evaluation of the potential benefits associated with introducing covariates are particularly important. To these purposes, we apply three testing approaches: the tests proposed by \cite{Be01} and \cite{AmGi07} and a loss-functions-based forecast evaluation that builds on the \cite{DiMa02} testing approach.

The first step of the \cite{Be01} test consists of computing, for all of the available days, the variable
$\nu_{t}=\int_{-\infty}^{fpc_{t}}\hat{f}_{t-1}(u) du = \hat{F}_{t-1}(fpc_{t})$,
where $\hat{f}_{t-1}(\cdot)$ is the \textit{ex ante} predicted density obtained at day $t-1$, and $fpc_{t}$ is the \textit{ex post} value of $FPC$ recorded at day $t$. Under correct model specification, \cite{Ro52} showed that $\nu_t$ is i.i.d. and uniformly distributed on $(0,1)$, a result that holds regardless of the underlying distribution of $fpc_t$, even when $\hat{F}_{t-1}(\cdot)$ changes over time. \cite{Be01} first pointed out that, if $\nu_t \sim \mathcal{U}(0,1)$, then $z_t=\Phi^{-1}(\nu_t) \sim \mathcal{N}(0,1)$, where $\Phi^{-1}(\cdot)$ denotes the inverse of the standard normal distribution function. The idea of the proposed 
test statistic, $LR_b$, is that, under a correct model specification, $z_t$ should be independent and identically distributed as standard normal; therefore an alternative hypothesis is that the mean and the variance differ from 0 and 1, respectively, with a first-order autoregressive structure. 

The \cite{Be01} test can be applied to models that provide a density forecast for the realized range volatility. The alternative models’ specifications for the conditional quantiles (such as with/without the covariates) or density forecast approaches may differ. Obviously, models that do not provide a rejection of the null hypothesis will be correctly specified, so, at least in principle, many alternative specifications could be appropriate for the data at hand.

The approach Berkowitz proposed allows for an absolute assessment of a given model. In fact, it focuses on the goodness of a specific sequence of density forecasts, relative to the unknown data-generating process. However, the Berkowitz test has a limitation in that it has power only with respect to misspecification of the first two moments. As \cite{Be01} noted, if the first two conditional moments are specified correctly, then the likelihood function is maximized at the conditional moments’ true values. Nevertheless, in practice, models could be misspecified even at higher-order moments. In that case, a viable solution is to compare density forecasts, that is, to perform a relative comparison given a specific measure of accuracy. To cope with this issue, in addition to \cite{Be01}’s approach, we consider \cite{AmGi07}’s test and a similar loss function-based approach that uses the \cite{DiMa02} test statistics.

\cite{AmGi07} developed a formal out-of-sample test for ranking competing density forecasts that is valid under general conditions. The test is based on a widely adopted metric, the \textit{log-score}, which in our framework is equal to $\log \hat{f}_{t-1}(fpc_{t})$, where $\hat{f}_{t-1}(.)$ is the \textit{ex ante} predicted density generated by our model at day $t-1$, and $fpc_t$ is the \textit{ex post} value of $FPC$ recorded at day $t$. Instead, we denote $\hat{g}_{t-1}(.)$ as the predictive density obtained by a competing model. For the two sequences of density forecasts, we define the quantity

\begin{equation} \label{WLR}
WLR_t=w \left( fpc_t^{st}\right) \left[\log \hat{f}_{t-1}(fpc_{t}) -\log \hat{g}_{t-1}(fpc_{t}) \right],
\end{equation}  

where $fpc_t^{st}$ is the realization of $FPC$ at day $t$, standardized using the estimates of the unconditional mean and standard deviation computed from the same sample on which the density forecasts for $t$ are estimated, and $w\left( fpc_t^{st}\right)$ is the weight the forecaster arbitrarily chooses to emphasize particular regions of the distribution's support. After computing the quantities $WLR_t$ for all of the samples considered in the forecast evaluation, we compute the mean $\overline{WLR}=(T-m)^{-1}\sum_{t=m+1}^{T} WLR_t$, where $m$ is the window size adopted for the computation of density forecasts.\footnote{The test can be used in the presence of a rolling approach for the computation of density forecasts. The value $m$ indicates the size of the rolling window or the fact that time $t$ forecasts depend, at maximum, on the last $m$ data points.}

In order to test for the null hypothesis of equal performance, that is, $H_0: E\left[\overline{WLR}\right]=0$, against the alternative of a different predictive ability $H_1: E\left[\overline{WLR}\right]\neq 0$, \cite{AmGi07} proposed the use of a weighted likelihood ratio test:

\begin{equation} \label{AGTEST}
AG=\frac{\overline{WLR}}{\hat{\sigma}_{AG}/\sqrt{T-m}},
\end{equation}

where $\hat{\sigma}_{AG}^2$ is a heteroskedasticity- and autocorrelation-consistent (HAC) \cite{NeWe87} estimator of the asymptotic variance $\sigma_{AG}^2=Var\left[\sqrt{T-m} \; \overline{WLR} \right]$. \cite{AmGi07} showed that, under the null hypothesis, $AG\stackrel{d}{\rightarrow}\mathcal{N} (0,1)$.

We applied the \cite{AmGi07} test by using five designs for the weights in equation (\ref{WLR}), which allows us to verify how the results change according to the particular regions of the distribution's support on which we are focusing. We set $w_{CE} \left( fpc_t^{st}\right)=\phi\left(fpc_t^{st} \right)$  when we focus on the center of the distribution, $w_{TL} \left( fpc_t^{st}\right)=1-\phi\left(fpc_t^{st} \right)/\phi(0)$ to concentrate on both the tails of the distribution, $w_{RT} \left( fpc_t^{st}\right)=\Phi\left(fpc_t^{st} \right)$ for the right tail, and $w_{LT} \left( fpc_t^{st}\right)=1-\Phi\left(fpc_t^{st} \right)$ for the left tail, and $w_{NW} \left( fpc_t^{st}\right)=1$ when giving equal importance to the entire support ($\phi(.)$ and $\Phi(.)$ denote the standard normal density function and the standard normal distribution function, respectively). 

Finally, we carried out a comparison at the single quantile level, focusing on the quantiles that have critical importance in our framework. To this purpose, we built on the approach \cite{DiMa02} proposed and considered the following loss function:

\begin{eqnarray}
L_{\tau,t}^{(i)}\left(fpc_t,Q_{fpc_t}^{(i)}(\tau,\textbf{x}_{t-1})\right) & = & \left[\tau-I \left( fpc_t-Q_{fpc_t}^{(i)}(\tau,\textbf{x}_{t-1})<0 \right)   \right] \nonumber \\ & \times & \left(fpc_t- Q_{fpc_t}^{(i)}(\tau,\textbf{x}_{t-1}) \right),
\label{LossFuncQ}
\end{eqnarray}

where $Q_{fpc_t}^{(i)}(\tau,\textbf{x}_{t-1})$ is the $\tau$-th forecasted quantile of $FPC$, obtained from the $i$-th model.

Let $d_{DM,\tau,t}$ be the loss differential between the quantile forecasts from two competitive models, $i$  and $j$ (where $i$ represents our proposal), that is, $d_{DM,\tau,t}=L_{\tau,t}^{(i)}\left(fpc_t,Q_{fpc_t}^{(i)}(\tau,\textbf{x}_{t-1})\right)-L_{\tau,t}^{(j)}\left(fpc_t,Q_{fpc_t}^{(j)}(\tau,\textbf{x}_{t-1})\right)$. After computing the quantities $d_{DM,\tau,t}$ for the forecasting sample, we compute the mean: $\overline{d_{DM,\tau}}=(T-m)^{-1}\sum_{t=m+1}^{T} d_{DM,\tau,t}$. We are interested in testing the null hypothesis $H_0: E\left[\overline{d_{DM,\tau}}\right]=0$ against the alternative $H_1: E\left[\overline{d_{DM,\tau}}\right]\neq 0$. To that purpose, we compute the \cite{DiMa02} test for $\tau=\{0.1,0.5,0.9\}$ and the associated test statistics are denoted, respectively, by $DM_{0.1}$, $DM_{0.5}$ and $DM_{0.9}$.

\section{Results}

\subsection{Full sample analyses}
First, we focus on model (\ref{mainmodel}) to analyze the full-sample estimated parameters and their p-values. This model was built to analyze and forecast the conditional quantiles of the first principal component of the realized range-based bias corrected bipower variations. For simplicity, Table \ref{quant_mainmult} reports just the results associated with $\tau=\{ 0.1,0.5,0.9\}$. The standard errors are computed by means of a bootstrapping procedure using the $xy$-pair method, which provides accurate results without assuming any particular distribution for the error term. 

When $\tau$ equals 0.1, only $sp500_{t-1}$ is not significant at the 95\% level. All the coefficients have small p-values at $\tau=\{0.2,0.3,0.4,0.5\}$, while for $\tau=\{0.6,0.7\}$ only $jump_{t-1}$ is not significant at the 95\% level. Finally, when $\tau>0.7$, only $fpc_{t-1}$ and $sp500_{t-1}$ are highly significant. Therefore, the first important result is that the variables that significantly affect $Q_{fpc_t}(\tau|\mathbf{x}_{t-1})$ change according to the $\tau$ level. Notably, only $fpc_{t-1}$ is always significant, whereas $sp500_{t-1}$ is not significant only when $\tau=0.1$. It is important to highlight that only $fpc_{t-1}$ and $sp500_{t-1}$ are significant in order to explain the high quantiles of volatility, which assume critical importance in finance. Moreover, the fact that $jump_{t-1}$ is not significant for high values of $\tau$ is a reasonable result since the volatility is already in a \textquotedblleft high \textquotedblright state, and we might safely assume that the jump-risk is already incorporated in it.

\begin{table}[htb]
\begin{center}
\centering
\small
\captionof{table}{Quantile regression results.} 
\begin{tabular}{ccc}
\hline
Variable & Coefficient Value & P-value \\
\hline
\multicolumn{3}{c}{$\tau=0.1$} \\
\hline
$fpc_{t-1}$ & 0.24157 & 0.00548 \\
$\overline{fpc_5}_{t-1}$ & 0.15385 & 0.03927 \\
$vix_{t-1}$ & 0.00020 & 0.00001 \\
$sp500_{t-1}$ & -0.00119 & 0.17431 \\
$jump_{t-1}$ & 0.87638 & 0.03242 \\
\hline
\multicolumn{3}{c}{$\tau=0.5$} \\
\hline
$fpc_{t-1}$ & 0.44580 & 0.00000 \\
$\overline{fpc_5}_{t-1}$ & 0.28779 & 0.00013 \\
$vix_{t-1}$ & 0.00019 & 0.00006 \\
$sp500_{t-1}$ & -0.00339 & 0.00002 \\
$jump_{t-1}$ & 0.77113 & 0.01017 \\
\hline
\multicolumn{3}{c}{$\tau=0.9$} \\
\hline
$fpc_{t-1}$ & 1.44195 & 0.00000 \\
$\overline{fpc_5}_{t-1}$ & 0.15027 & 0.52722 \\
$vix_{t-1}$ & -0.00002 & 0.87801 \\
$sp500_{t-1}$ & -0.00919 & 0.00000 \\
$jump_{t-1}$ & -0.07240 & 0.90576 \\
\hline
\end{tabular}\par
\label{quant_mainmult} 
\end{center}
\footnotesize{The table reports the coefficients and the p-values for model (\ref{mainmodel}); $\tau=\{0.1,0.5,0.9\}$. The standard errors are computed by means of the bootstrapping procedure, by employing the $xy$-pair method.}
\end{table}

These conclusions are consistent with the values of the pseudo-coefficient of determination, $R^1(\tau)$, and the test statistic $\xi_w$ proposed by \cite{KoBa82b}.\footnote{See Appendix B for more details.} We note that $R^1(\tau)$ is a local measure of fit, so it can't be used for a global goodness-of-fit assessment. However, it is possible to compare a restricted model that has fewer explanatory variables against an unrestricted one that includes all of the variables under consideration. The unrestricted model is defined by equation (\ref{mainmodel}). Moreover, in order to link the conclusions that arise from the various quantile regressions considered with those analyzed above, the restricted model includes only the most relevant explanatory variables: $fpc_{t-1}$ and $sp500_{t-1}$. 

\begin{table}[htb]
\begin{center}
\centering
\small
\captionof{table}{Restricted model against unrestricted model.} 
\begin{tabular}{cccc}
\hline
$\tau$ 	& $R^1(\tau)$  			& $R^1(\tau)$ 		& $\xi_w$\\
		& Unrestricted model 	& Restricted model	& P-value\\
\hline
0.1 & 0.3655 & 0.2940 & 0.0000\\
0.2 & 0.4439 & 0.3785 & 0.0000\\
0.3 & 0.5005 & 0.4422 & 0.0000\\
0.4 & 0.5427 & 0.4952 & 0.0000\\
0.5 & 0.5801 & 0.5396 & 0.0000\\
0.6 & 0.6142 & 0.5803 & 0.0000\\
0.7 & 0.6474 & 0.6223 & 0.0000\\
0.8 & 0.6865 & 0.6755 & 0.0007\\
0.9 & 0.7492 & 0.7467 & 0.9123\\
\hline
\end{tabular}\par
\label{pseudo_coeff} 
\end{center}
\footnotesize{The table reports some results coming from the comparison between the unrestricted model, that is model (\ref{mainmodel}), and the restricted model, in which the regressors are just $fpc_{t-1}$ and $sp500_{t-1}$. The first two columns give the pseudo coefficients of determination, whereas the third one gives the p-values of the test statistic $\xi_w$.}
\end{table}

Table \ref{pseudo_coeff} shows the values of the pseudo-coefficient of determination computed for the restricted and unrestricted models at $\tau= \{$0.1, 0.2, 0.3, 0.4, 0.5, 0.6, 0.7, 0.8, $0.9\}$. We first observe that $R^1(\tau)$ is a positive function of $\tau$ for both the restricted and unrestricted models and then note that the differences between the restricted and unrestricted models decrease as $\tau$ increases until the differences all but disappear at $\tau=0.9$. Therefore, the contribution of $\overline{fpc_5}_{t-1}$, $vix_{t-1}$, and $jump_{t-1}$ to the goodness-of-fit of model (\ref{mainmodel}) is largely irrelevant at $\tau=0.9$. We obtain similar conclusions from the test proposed in \cite{KoBa82b}. The null hypothesis is that the additional variables used in the unrestricted model do not significantly improve the goodness-of-fit with respect to the restricted model. Table \ref{pseudo_coeff} shows the p-values of the test statistic $\xi_w$: the null hypothesis of the test is accepted at $\tau=0.9$, while at lower quantiles, some of the additional variables provide sensible improvements in the model fit.

Returning to the coefficients’ values, we note that their impact on the conditional quantiles of $fpc_{t-1}$, $\overline{fpc_5}_{t-1}$, $vix_{t-1}$, and $jump_{t-1}$ is positive in all cases in which the coefficients are statistically significant. Therefore, there is a positive relationship between these variables and $Q_{fpc_t}(\tau|\mathbf{x}_{t-1})$. With respect to the price jumps, the positive impact is a somewhat expected result, and it extends \cite{CoPiRe10}’s findings on prices jumps’ impact on realized volatility. However, since the coefficient of $sp500_{t-1}$, $\hat{\beta}_4(\tau)$, is always negative, in keeping with \cite{Bl76}, we find that an increasing market return implies greater stability and negative effects on $Q_{fpc_t}(\tau|\mathbf{x}_{t-1})$. We also checked on the persistence of volatility, measured by the sum of the HAR coefficients, that is, $\hat{\beta}_1(\tau)+\hat{\beta}_2(\tau)$, and noted that persistence is stronger at high levels of $\tau$: $\hat{\beta}_1(0.1)+\hat{\beta}_2(0.1)= 0.396$, $\hat{\beta}_1(0.5)+\hat{\beta}_2(0.5)=0.734$, and $\hat{\beta}_1(0.9)+\hat{\beta}_2(0.9)= 1.5923$. This result, which is coherent with the result on jumps, suggests that volatility in high regimes (upper quantiles) is more persistent as opposed to median or low regimes (lower quantiles); in addition, that unexpected movements/shocks (including jumps) may have a larger effect on lower volatility quantiles compared to their impact on higher volatility quantiles, as they convey relevant information. While these results, recovered from a full-sample analysis, provide an interesting interpretation, they do not take into account the possible structural changes in the relationship between covariates and volatility conditional quantiles. This problem is analyzed below. 

\begin{figure}[htb]
\centering
\includegraphics[scale=0.85]{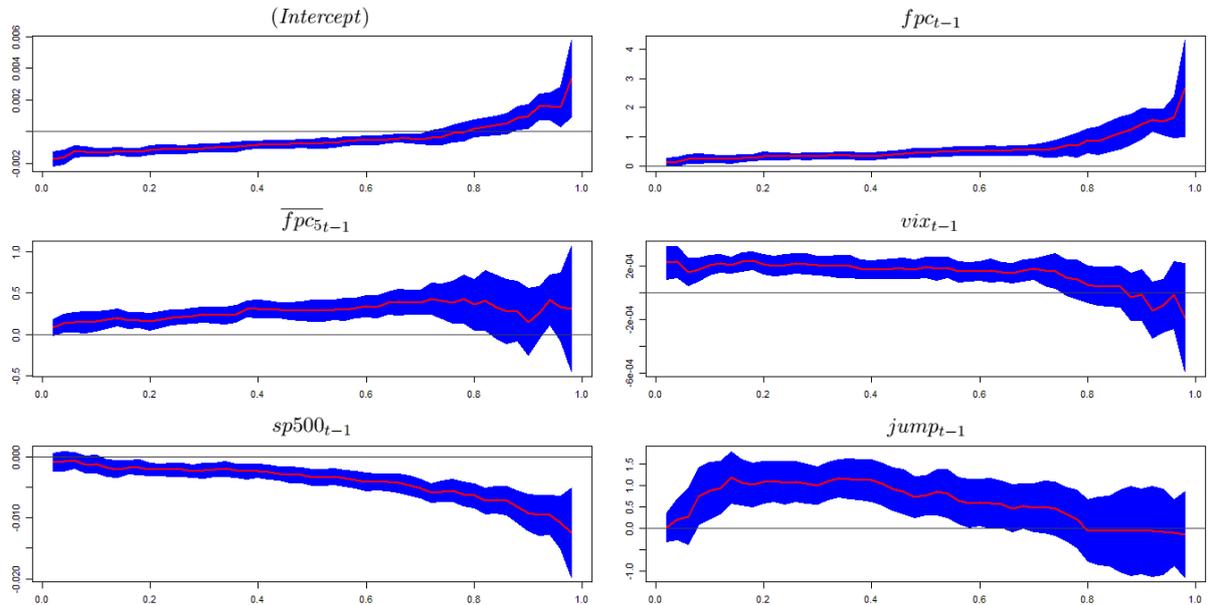}
\caption{Coefficients plots. The red lines represent the coefficients values over $\tau$ levels, while the blue areas are the associated 95\% confidence intervals.}
\label{coef_plot}
\end{figure}

Even if the coefficients’ signs don't change over $\tau$, it's important to determine whether changes in $\tau$ affect their magnitude. In other words, we want to check the so-called location-shift hypothesis, which states that the parameters in the conditional quantile equation are identical over $\tau$. Important information can be drawn from Figure \ref{coef_plot}, which provides the coefficients’ plots.

The impact of $fpc_{t-1}$ on the conditional quantiles of volatility is constant up to $\tau=0.7$, where it increases significantly. In the case of $\overline{fpc_5}_{t-1}$, we observe a slightly increasing trend until $\tau=0.7$, when the uncertainty level becomes noticeable. The impact of $vix_{t-1}$ has a flat trend up to $\tau=0.7$, when it begins a decreasing trend, reaching negative values in a region where the regressor is not significant. However, $sp500_{t-1}$ shows a negative effect on the volatility quantiles, and this relationship grows quickly at high values of $\tau$. We associate this finding with the so-called \textit{leverage effect}, as argued by \cite{Bl76}: increases in volatility are larger when previous returns are negative than when they have the same magnitude but are positive. To verify this claim, we divided the $fpc_t$ series into deciles and, conditioning to those deciles, computed the mean of $sp500_{t-1}$ for the various groups. As expected, the mean of $sp500_{t-1}$, corresponding to the values of $fpc_t$ in the first decile, is 0.19\%, whereas the mean corresponding to the last decile, in which we have the highest $fpc_t$ values, is -0.28\%. As a consequence, the negative coefficients for $sp500_{t-1}$ can be seen as supporting the existence of the leverage effect. Finally, in the case of $jump_{t-1}$, we observe a wide band, where its impact grows at the beginning but takes negative turns from $\tau=0.4$. 

To summarize, we verify that the relationships between the regressors and the response variable are not constant over $\tau$. In particular, the impact of $fpc_{t-1}$ and $sp500_{t-1}$ grows considerably at high values of $\tau$. Therefore, $fpc_{t-1}$ and $sp500_{t-1}$ are critical indicators in the context of extreme events where volatility can reach high levels. The coefficients of the other explanatory variables do not exhibit particular trends at low-medium levels of $\tau$, and when $\tau$ assumes high values, they become even more volatile, in a region of high uncertainty, given their wide confidence bands. 

Analysis of the coefficients’ plots shown in Figure \ref{coef_plot} suggests that the hypothesis of equal slopes does not hold. In order to reach more accurate conclusions, we perform a variant of the Wald test introduced by \cite{KoBa82}. The null hypothesis of the test is that the coefficient slopes are the same across quantiles. The test is performed taking into account three distant values of $\tau$, $\tau=\{0.1,\;0.5,\;0.9\}$, to cover a wide interval. When we compare the models estimated for $\tau=0.1$ and $\tau=0.5$, the null hypothesis is rejected at the 95\% confidence level of for $fpc_{t-1}$, $\overline{fpc_5}_{t-1}$, and $sp500_{t-1}$. When we consider $\tau=0.5$ and $\tau=0.9$, the null hypothesis is rejected at the 99\% confidence level for $fpc_{t-1}$ and $sp500_{t-1}$. We obtain the same result when we focus on $\tau=0.1$ and $\tau=0.9$, so we have evidence against the location shift hypothesis for those regressors. This finding confirms that the relationship between covariates and conditional quantiles varies across quantile values. This fundamentally relevant finding highlights that, when the interest lies on specific volatility quantiles, linear models can lead to inappropriate conclusions on whether there is a relationship between covariates and volatility measures, and if there is, on the strength of the relationship.

\subsection{Rolling analysis}

The U.S. subprime crisis and the European sovereign debt crisis have had noticeable effects on the financial system, with possible impacts also on the relationship between volatility and its determinants. Therefore, it is important to determine whether these events also affect the parameters of model (\ref{mainmodel}). To this end, we perform a rolling analysis with steps of one day, using a window size of 500 observations and $\tau$ ranging from 0.05 to 0.95 with steps of 0.05. Thus, we consider nineteen levels of $\tau$ and, for a given $\tau$, obtain 2,133 estimates of a single coefficient. The finer grid adopted here allows us to recover a more accurate picture of the evolution of conditional quantiles. Nevertheless, the most relevant quantiles in this case are the upper quantiles, which are associated with the highest volatility levels. The estimated coefficients across time and quantiles are summarized in several figures.

\begin{figure}[htb]
\centering
\includegraphics[scale=0.85]{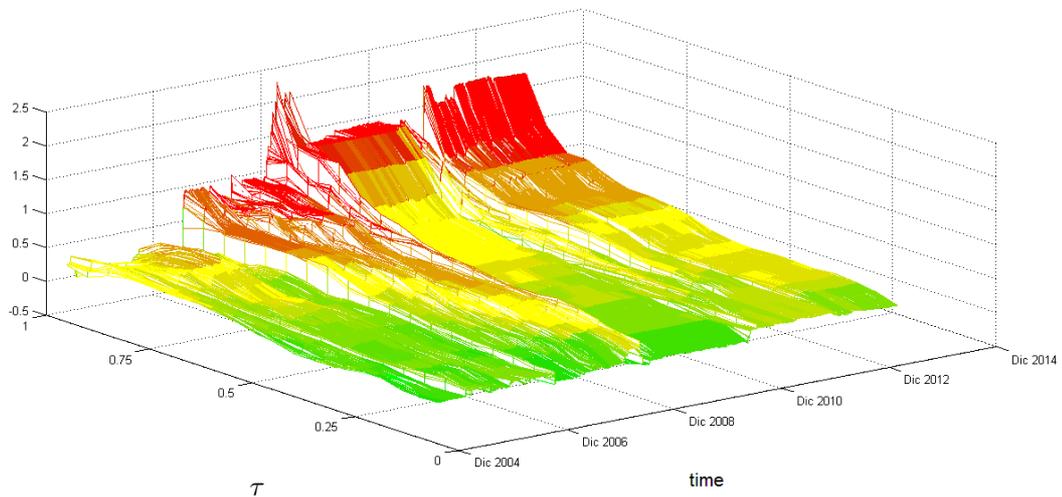}
\caption{Rolling analysis for $fpc_{t-1}$.}
\label{roll_lag_FPC}
\end{figure}

Figure \ref{roll_lag_FPC} reports the evolution of the relationship between $fpc_{t-1}$ and the conditional volatility quantiles over time and over $\tau$. The first result that arises from Figure \ref{roll_lag_FPC} is that the impact of $fpc_{t-1}$ has a comparatively stable trend over time for medium-low $\tau$ levels; some jumps are recorded, mainly in the period of the subprime crisis, but their magnitude is negligible. The picture significantly changes in the region of high $\tau$ levels, where the surface is relatively flat and lies at low values in the beginning, but after the second half of 2007, when the effects of the subprime crisis start to be felt, there is a clear increase in the coefficient values, which reach their peak in the months between late 2008 and the beginning of 2009. Moreover, in this period we record the highest volatilities in the $fpc_{t-1}$ coefficients over $\tau$ levels. In the following months, the coefficient values decrease, but they remain at high levels until the end of the sample period. $fpc_{t-1}$ is a highly relevant variable for explaining the entire conditional distribution of $fpc_{t}$ since it is statistically significant over a large number of quantiles. Figure \ref{roll_lag_FPC} verifies that the relationship between $fpc_{t}$ and $fpc_{t-1}$ is affected by particular events, such as the subprime crisis, mainly at medium-high $\tau$ levels. This finding confirms the change in the parameter across $\tau$ values, with an increasing pattern in $\tau$ and highlights that, during periods of market turbulence where the volatility stays at high levels, the volatility density overreacts to past movements of volatility, since the $fpc_{t-1}$ coefficient is larger than $1$ for upper quantiles. Therefore, after a sudden increase in volatility at, say, time $t$, we have an increase in the conditional quantiles for time $t+1$ and, therefore, an increase in the likelihood that we will observe additional volatility spikes (that is, volatility that exceeds a time-invariant threshold) at time $t+1$.

\begin{figure}[htb]
\centering
\includegraphics[scale=0.85]{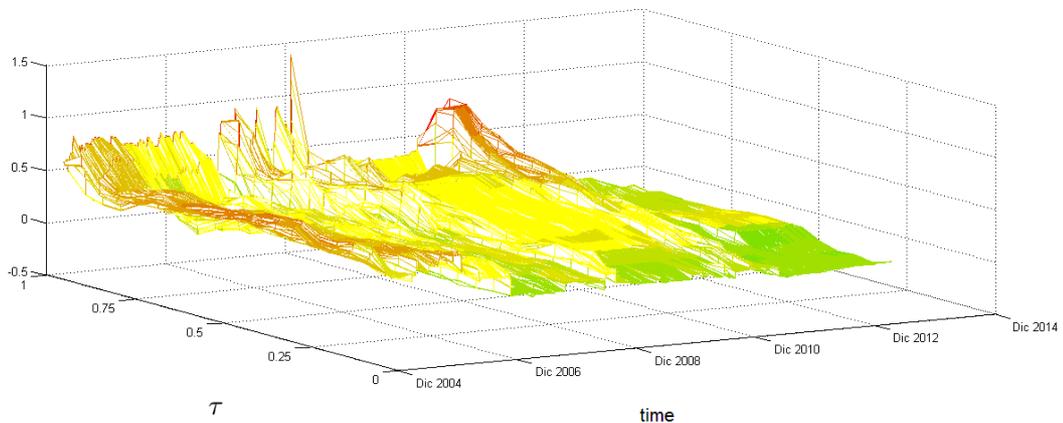}
\caption{Rolling analysis for $\overline{fpc_5}_{t-1}$.}
\label{roll_lag_FPC_5}
\end{figure}

Referring to $\overline{fpc_5}_{t-1}$, the HAR coefficient reported in Figure \ref{roll_lag_FPC_5} has a volatile pattern until late 2008, when it reaches its  peak. After that, the surface flattens, but another jump is recorded in mid-2011, mainly in the region of high $\tau$ values. Therefore, the relationship between the $fpc_t$ quantiles and $\overline{fpc_5}_{t-1}$, which reflects the perspectives of investors who have medium time horizons, is volatile over time, mainly in the region of high $\tau$ values. Again, this result can be associated with crises that affect the persistence and the probability that extreme volatilities will occur.

Section 5.1 pointed out that the persistence of volatility, measured by the sum of the HAR coefficients ($\hat{\beta}_1(\tau)+\hat{\beta}_2(\tau)$), is stronger at high levels of $\tau$ than it is at lower levels. Using the rolling analysis, we also determined how that persistence evolves over time. We focused on 
$\tau=\{0.1,0.5,0.9\}$, observing that the persistence is always positive, as one might expect, and that it has relevant differences across quantiles. Looking over time, we confirm the full-sample result that persistence increases with $\tau$ levels and note that the reaction of the persistence to the subprime crises is clear in all three cases but is most pronounced for $\tau=0.9$. However, the European sovereign debt crisis affects only the $\tau=0.9$ case, where we note an increase in persistence in the last part of the sample.

\begin{figure}[htb]
\centering
\includegraphics[scale=0.85]{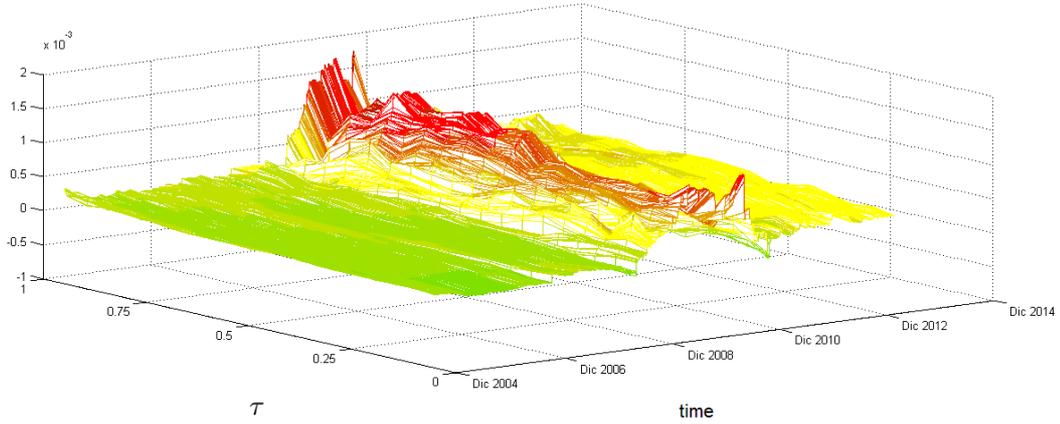}
\caption{Rolling analysis for $vix_{t-1}$.}
\label{roll_lag_VIX}
\end{figure}

Figure \ref{roll_lag_VIX} shows two periods in which the $vix_{t-1}$ coefficient has high values: between the end of 2008 and early 2010 and a shorter period from the end of 2011 to the first half of 2012. While in the first period the impact of $vix_{t-1}$ significantly increases for all $\tau$ levels, in the second period the increase in the coefficient affects just the surface region in which $\tau$ takes high values. Unlike the HAR coefficients described above, the $vix_{t-1}$ coefficient does not have a clear and stable increasing trend over $\tau$ levels. In addition, the relationship between the $fpc_t$ conditional quantiles and $vix_{t-1}$ is highly sensitive to the subprime crisis, when pessimism among financial operators, reflected in the implied volatility of the S\&P 500 index options, was acute. This result is again somewhat expected since we focus on U.S.-based data and the subprime crisis had a high impact on the U.S. equity market. Our results are evidence that the perception of market risk has a great impact on the evolution of market volatility (as proxied by $fpc_t$), particularly during financial turmoil. The impact is not so clear-cut during the European sovereign crisis, which had less effect on the U.S. equity market.
 
\begin{figure}[htb]
\centering
\includegraphics[scale=0.85]{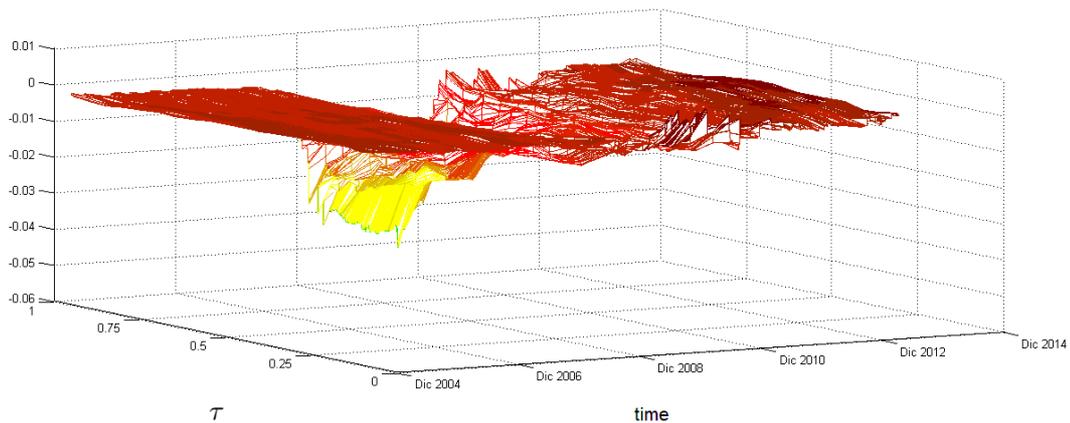}
\caption{Rolling analysis for $sp500_{t-1}$.}
\label{roll_lag_SP500}
\end{figure}

Figure \ref{roll_lag_SP500} shows how the impact of $sp500_{t-1}$ evolves over time and over $\tau$. The surface given is almost always flat, the exception being the months between late 2008 and the end of 2010, when the effects of the subprime crisis were particularly acute; during this time the coefficient values decrease as $\tau$ grows, mainly for values of $\tau$ above the median. The lagged value of the S\&P 500 index return affects the entire conditional distribution of $fpc_t$ and is statistically significant in almost all of the quantiles considered. Moreover, Figure \ref{roll_lag_SP500} shows that the effect is negative and particularly pronounced during the subprime crisis, when negative returns exacerbated market risk, increasing the upper quantiles’ volatility and increasing the likelihood of large and extreme volatility events.

\begin{figure}[htb]
\centering
\includegraphics[scale=0.85]{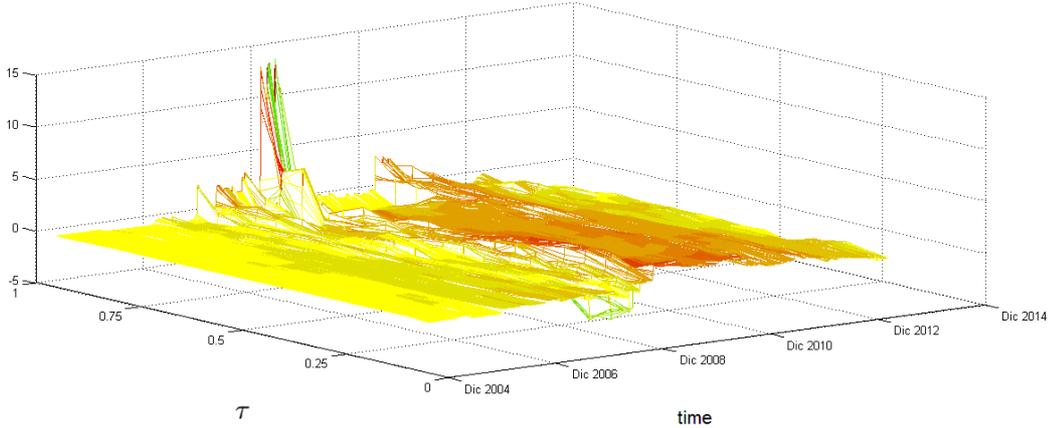}
\caption{Rolling analysis for $jump_{t-1}$.}
\label{roll_lag_jump}
\end{figure}

Finally, Figure \ref{roll_lag_jump} reports the surface associated with the price jump. At the beginning of the sample, the $jump_{t-1}$ coefficient takes on small values over the $\tau$ levels; however, it starts to grow in 2007, reaching high peaks at high $\tau$ levels during the subprime crisis. Although the coefficient reaches considerable values in this region, their statistical significance is limited. After the second half of 2009, the surface flattens out again until the end of the sample, with the exception of some peaks of moderate size that were recorded in 2011 during the sovereign debt crises. 

To summarize, using the rolling analysis, we show that two special and extreme market events (the U.S. subprime crisis and the European sovereign debt crisis) affected the relationships between the realized volatility quantiles and a set of covariates. Our results show that coefficients can reasonably vary, with a potential and relevant impact on the forecasts of both the mean (or median) volatility and the volatility distribution (starting from the quantiles). The effects differ across quantiles and change with respect to the volatility upper tails, as compared to the median and the lower tail. Therefore, when volatility quantiles are modeled, the impacts of covariates might differ over time and over quantiles, being crucial during certain market phases. This result further supports the need for quantile-specific estimations when there is an interest in single volatility quantiles.

\subsection{Evaluation of the predictive power}

We evaluate the volatility density forecasts by means of the tests by \cite{Be01}, \cite{AmGi07}, and \cite{DiMa02} proposed, the details of which are given in Section 4. The Berkowitz test is carried out by estimating more $fpc_t$ conditional quantiles with respect to the analyses discussed in Section 5.2 in order to have smoother distribution functions. In particular, we consider forty-nine values of $\tau$, ranging from 0.02 to 0.98, with steps of 0.02. Given the findings of the previous subsection, we must use a rolling procedure, but we modify the rolling scheme previously used to keep a balance between the reliability of the estimated coefficients and computational times. 

The Berkowitz test is applied on the volatility density forecasts recovered based on conditional quantiles estimated from subsamples of 100 observations with steps of ten days. The estimated quantiles are then linearly interpolated to recover the entire volatility distribution. Thus, the conditional distribution of $fpc_t$ is estimated for each subsample and is linked to the corresponding out-of-sample observation, and we obtain the $z_t$ series. In the case of model (\ref{mainmodel}), $z_t$ is normally distributed, as the likelihood ratio test $LR_b$ equals 5.26 and the null hypothesis of the Berkowitz test, that is, $z_t \sim \mathcal{N}(0,1)$ with no autocorrelation, is not rejected at the 95\% level, validating the forecast goodness of the model (\ref{mainmodel}). 

To determine whether the predictive power of our approach is affected by the U.S. subprime crisis and the European sovereign debt crisis, the series $z_t$ is divided into two parts of equal length: the first referring to a period of relative calm from the beginning of 2003 to the first half of 2007 and the second referring to a period of market turmoil that was due to the two crises, between the second half of 2007 and the first half of 2013. In the first part $LR_b$ equals 2.34, and in the second it equals 5.39. Nevertheless, the null hypothesis of the Berkowitz test is not rejected at the 95\% level in both the cases. Therefore, the conditional quantile model and the approach we adopt to recover the conditional density forecasts are appropriate even during financial turbulence.

Analysis of the results reveals that, as in the analysis of the full sample, $\overline{fpc_5}_{t-1}$,  $vix_{t-1}$, and $jump_{t-1}$ are not sufficient to explain the volatility quantiles at high values of $\tau$ in many of the subsamples. Therefore, we must determine whether this result affects the output of the Berkowitz test using a restricted model in which the regressors are only $fpc_{t-1}$ and $sp500_{t-1}$. $LR_b$ equals 2.60, a smaller value than the previous cases. The last findings reported above suggest that the inclusion of non-significant explanatory variables penalizes the predictive power of model (\ref{mainmodel}). The restricted model gives the lowest value of $LR_b$, but the predictive power could be improved by selecting only those variables that are significant for each value of $\tau$ and for each subsample in order to forecast the conditional distribution of the volatility. Thus, the structure of model (\ref{mainmodel}) would change over time and over $\tau$. However, this approach is not applied in the present work since it would require using only the significant variables in forty-nine models, one for each specific value of $\tau$, while it should be considered across all of the rolling subsamples.

We have focused thus far on an absolute assessment of our approach. Now we compare it with a competing model that is fully parametric. We recover the predictive conditional distribution of the realized range volatility by means of a HARX model in which the mean dynamic is driven by a linear combination of the explanatory variables $fpc_{t-1}$ and $\overline{fpc_5}_{t-1}$ (the HAR terms) and the exogenous variables $vix_{t-1}$, $sp500_{t-1}$, and $jump_{t-1}$ (the X in the model’s acronym). In addition, to capture the volatility-of-volatility effect of \cite{CoMiPiPi08}, a GJR-GARCH term \citep{GlJaRu93} is introduced on the innovation. The error term is also assumed to follow a normal-inverse Gaussian ($\mathcal{NIG}$) distribution with mean 0 and variance 1. Thus, the conditional variance is allowed to change over time, and the distribution of the error is flexible to features like fat tails and skewness. 

We start by using the Berkowitz test to evaluate the density forecast performance of the HARX-GJR model. The likelihood ratio test $LR_b$ equals 113.88, a high value that suggests a clear rejection of the null hypothesis. We also determined whether the HARX-GJR model works better when we use the logarithm of the volatility as a response variable, following the evidence in \cite{CoPiRe10}, and found that the likelihood ratio test $LR_b$ provides a much lower value (20.27). Even so, the test signals a rejection of the null hypothesis with a low p-value. As a first finding, our approach, based on the interpolation of multiple quantile forecasts, provides more flexibility than the parametric HARX-GJR model does and is better in terms of the Berkowitz test.

To provide more accurate results, we move to a comparison of our approach with the HARX-GJR by means of the \cite{AmGi07} and \cite{DiMa02} test. With regard to the \cite{AmGi07} test, we use five weights to compute the quantity given in equation (\ref{AGTEST}): $w_{CE} \left( fpc_t^{st}\right)$, $w_{TL} \left( fpc_t^{st}\right)$, $w_{RT} \left( fpc_t^{st}\right)$, $w_{LT} \left( fpc_t^{st}\right)$, and $w_{NW} \left( fpc_t^{st}\right)$. The associated likelihood ratio tests are denoted $AG_{CE}$, $AG_{TL}$, $AG_{RT}$, $AG_{LT}$, and $AG_{NW}$, respectively. Our approach provides better results overall than the HARX-GJR since $\overline{WLR}$ is always positive. However, the differences between the two models are not always statistically significant. In fact, $AG_{CE}=1.08$, $AG_{TL}=2.11$, $AG_{RT}=2.37$, $AG_{LT}=0.40$, and $AG_{NW}=1.75$, with  p-values of 0.280, 0.034, 0.017, 0.689, and 0.080, respectively. Therefore, the null hypothesis of equal performance is rejected at the 5\% level in the cases of $AG_{TL}$ and $AG_{RT}$ and at the 10\% level in the case of $AG_{NW}$. Notably, the two density-forecast approaches are statistically different when the right tail is the focus.

Similar results are obtained when we consider the single quantile loss function and the \cite{DiMa02} test statistic. Our approach provides lower losses overall since $\overline{d_{DM,\tau}}$ is negative for all of the levels of $\tau$ we considered ($0.1,0.5,0.9$). However, the differences are statistically significant only in the case of $\tau=0.9$, given that $DM_{0.1}=-1.173$ (0.241), $DM_{0.5}=-1.076$ (0.282), and $DM_{0.9}=-4.928$ (8.31e-07). 

In summary, the results we reported here demonstrate the good performance of our model, particularly when we focus on the right tail of the volatility distribution. The right tail assumes critical importance in our framework, as it represents periods of high risk in the market.

Our findings indicate another relevant contribution of this study, as the quantile regression approach we propose can be used to recover density forecasts for a realized volatility measure. These forecasts improve on those of a traditional approach because of the inclusion of quantile-specific coefficients. This feature of our approach might become particularly relevant in all empirical applications where predictive volatility density is required.

\subsection{Single asset results}

The results in Subsections 5.1-5.3 were based on a summary of the sixteen asset volatility movements, which was itself based on the first principal component. Now we search for confirmation of the main findings of model (\ref{mainmodel}) by running model (\ref{single_asset_model}) at the single-asset level and then for all sixteen assets. For simplicity, we applied model (\ref{single_asset_model}) just at 
$\tau=\{0.1,0.5,0.9\}$.

We focus first on the relationships between $Q_{rrv_{i,t}}(\tau|\mathbf{x}_{i,t-1})$, and $rrv_{i,t-1}$ for $i=1,...,16$. When $\tau$ equals 0.1, $rrv_{i,t-1}$ is not significant ($\alpha=0.05$) to explain the conditional volatility quantiles of eight assets: $ATT$, $CAT$, $HON$, $IBM$, $PEP$, $PRG$, $TWX$, and $TXN$. At $\tau=0.5$, $rrv_{i,t-1}$ is not significant only for $PRG$, and at $\tau=0.9$, $rrv_{i,t-1}$ is not significant for $PEP$ and $PRG$. Compared with the other regressors and in line with the results obtained for the first principal component, $rrv_{i,t-1}$ is one of the most significant explanatory variable at high levels of $\tau$, so it assumes critical importance in the context of extreme events. The $rrv_{i,t-1}$ coefficient takes a negative value only for $PEP$ at $\tau=0.1$, but here it is not statistically significant. In all the other cases, it is always positive. Moreover, the magnitude of the impact that $rrv_{i,t-1}$ has on $Q_{rrv_{i,t}}(\tau|\mathbf{x}_{i,t-1})$ is a positive function of $\tau$ for all sixteen assets. The differences among the assets increase as $\tau$ grows, and at $\tau=0.9$ the financial companies ($BAC$, $CTG$, $JPM$ and $WFC$) record the highest coefficient values, highlighting the crucial importance of the extreme events in the financial system. $fpc_{t-1}$, the homologous regressor included in model (\ref{mainmodel}), has a weaker impact on the conditional volatility quantiles than $rrv_{i,t-1}$ does only for the financial companies $BAC$ ($\tau=0.1$),  $JPM$ ($\tau=0.5$), and $CTG$ ($\tau=0.9$). Therefore, the relationships between the conditional volatility quantiles and the lagged value of the response variable are stronger for the first principal component than for the single assets.

$\overline{rrv_5}_{i,t-1}$ is not significant ($\alpha=0.05$) at $\tau=0.1$ for $CTG$, $JPM$, and $PRG$, as the p-values of its coefficient indicate. At $\tau=0.5$ it is not significant only for $PRG$, whereas when $\tau$ equals 0.9, it is not significant for $CTG$, $JPM$, and $PRG$. The $\overline{rrv_5}_{i,t-1}$ coefficient is always positive and, with the exception of $CTG$ and $JPM$, it is a positive function of $\tau$. Moreover, the differences among the $\overline{rrv_5}_{i,t-1}$ coefficient values are more marked at high $\tau$ levels for all sixteen assets. The p-values of the $vix_{t-1}$ coefficient are less than 0.05 at $\tau=\{ 0.1,0.5\}$ for all sixteen assets. When $\tau$ equals 0.9, $vix_{t-1}$ is significant in the cases of $HON$, $HPQ$, $PRG$, $TWX$ and $TXN$. Unlike the coefficients previously mentioned, that of $vix_{t-1}$ 
does not have a particular trend over $\tau$; it takes positive values for all of the assets, so, as in the context of the first principal component, it has a positive impact on the conditional volatility quantiles. In addition, at $\tau=\{0.1,0.5\}$, the $vix_{t-1}$ coefficient is larger in model (\ref{mainmodel}) than it is in model (\ref{single_asset_model}) for all sixteen companies. Therefore, $vix_{t-1}$ has a more marked impact on the conditional volatility quantiles of the first principal component. The comparisons made at $\tau=0.9$ are useless given the high p-values of the coefficients of interest.

$sp500_{t-1}$ is always significant ($\alpha=0.05$) for the sixteen assets, with the exception of $BAC$, $CTG$, and $JPM$ at $\tau=0.1$. Its coefficient is always negative, as expected, so $sp500_{t-1}$ has a negative impact on the conditional volatilities’ quantiles. In addition, the magnitude of the impact is a negative function of $\tau$ for all sixteen assets, and those relationships become more marked at high $\tau$ levels. With the exception of one case ($TXN$ at $\tau=0.1$), the impact of $sp500_{t-1}$ at $\tau=\{0.1,0.5,0.9\}$ is more pronounced on the conditional volatility quantiles of the first principal component than it is when the assets are considered individually, as comparing the absolute values of the related coefficients shows. 

At $\tau=0.1$, $jump_{i,t-1}$ is significant ($\alpha=0.05$) for $CTG$, $IBM$, and $TWX$. It is significant only for $CTG$ and $TWX$ at $\tau=0.5$, whereas it is never significant when $\tau$ equals 0.9. $jump_{i,t-1}$’s coefficient takes both negative and positive values and, with the exception of a few assets, it does not have a particular trend over $\tau$. Comparing these results with those obtained for the $fpc_t$ conditional quantiles, we find that, with the exception of $CTG$ ($\tau=0.1$) and $PRG$ ($\tau=0.5$), the lagged value of the jump component has more impact in model (\ref{mainmodel}) than in model (\ref{single_asset_model}) at $\tau=\{0.1,0.5\}$. It is pointless to compare the coefficients at $\tau=0.9$ given their high p-values.

To summarize, the explanatory variables $rrv_{i,t-1}$, $\overline{rrv_5}_{i,t-1}$, $vix_{t-1}$, and $sp500_{t-1}$ are sufficient to explain the conditional volatility quantiles of the sixteen assets in most of the cases studied. Their coefficients tend to take the same sign for the sixteen assets: positive in the cases of $rrv_{i,t-1}$, $\overline{rrv_5}_{i,t-1}$, and $vix_{t-1}$ and negative in the case of $sp500_{t-1}$. Moreover, the coefficients of $rrv_{i,t-1}$, $\overline{rrv_5}_{i,t-1}$, and $sp500_{t-1}$ have a clear trend over $\tau$, providing evidence against the location-shift hypothesis, which assumes homogeneous impacts of the regressors across quantiles. However, $jump_{i,t-1}$ is significant in only a few cases. Furthermore, with the exception of $\overline{rrv_5}_{i,t-1}$, we find that, in most of the cases studied, the relationships between the explanatory variables and the conditional volatility quantiles are more pronounced in the context of the first principal component than when the assets are considered individually. This result shows that the first principal component captures a kind of systematic effect, where the relationship between macro and finance covariates and volatility quantiles is clearer. At the single-asset level the impact of covariates is more heterogeneous than for the first principal component, perhaps suggesting the need for company- (or sector-) specific covariates.

\begin{table}[htb!]
\begin{center}
\centering
\small
\captionof{table}{Berkowitz test for the single asset analysis.} 
\begin{tabular}{ccc}
\hline
Asset & $LR_b$ & $LR_{b,HARX-GJR}$ \\
\hline
ATT & 10.23 (0.0167) & 150.26 (0.0000) \\
BAC & 3.00 (0.3916) &  125.11 (0.0000)\\
BOI & 7.90 (0.04812) & 93.09 (0.0000)\\
CAT & 9.31 (0.0254) & 88.11 (0.0000)\\
CTG & 4.28 (0.2327) & 142.99 (0.0000)\\
FDX & 12.62 (0.0055) & 95.63 (0.0000)\\
HON & 5.95 (0.1140) & 115.59 (0.0000)\\
HPQ & 5.33 (0.1491) & 83.80 (0.0000)\\
IBM & 4.25 (0.2357) & 157.06 (0.0000)\\
JPM & 3.26 (0.3532) & 124.01 (0.0000)\\
MDZ & 5.49 (0.1392) & 159.86 (0.0000)\\
PEP & 9.97 (0.0188) & 164.45 (0.0000)\\
PRG & 2.25 (0.5222) & 130.88 (0.0000)\\
TWX & 11.09 (0.0112) & 99.10 (0.0000)\\
TXN & 1.90 (0.5934) & 79.09 (0.0000)\\
WFC & 7.04 (0.0706) & 143.04 (0.0000)\\
\hline
\end{tabular}\par
\label{Berkowitz_single} 
\end{center}
\footnotesize{The table reports for each stock (the ticker is given in the first column) the values of the likelihood ratio test (the p-values are given in brackets) proposed by \cite{Be01}, generated by model (\ref{single_asset_model}), $LR_b$, and the HARX-GJR model, $LR_{b,HARX-GJR}.$}
\end{table}

The last point of our analysis refers to the assessment of model (\ref{single_asset_model})’s predictive power, which we apply for each of the sixteen assets. As in the case of the model for the first principal component, we use the tests \cite{Be01}, \cite{AmGi07}, and \cite{DiMa02} proposed. With regard to the Berkowitz test, Table \ref{Berkowitz_single} provides the values of the likelihood ratio $LR_b$ and the results generated by model (\ref{single_asset_model}), showing that the null hypothesis of the test, that is, $z_t \sim \mathcal{N}(0,1)$ with no autocorrelation, is not rejected for ten assets: $BAC$, $CTG$, $HON$, $HPQ$, $IBM$, $JPM$, $MDZ$, $PRG$, $TXN$, and $WFC$. The results from the other six cases stem from the fact that some variables, mainly $jump_{i,t-1}$, are not significant in many subsamples for several $\tau$ levels. As indicated in Section 5.3, which focused on the first principal component, the predictive power of our approach could be improved by selecting, for each subsample and each $\tau$, only the regressors that are significant in order to explain the individually evaluated conditional quantiles. Thus, the structure of model (\ref{single_asset_model}) would change over time. Now we compare our approach with the HARX-GJR model. The results that arise from using the HARX-GJR model are given in the third column of Table \ref{Berkowitz_single}. As in the first principal component context, the likelihood ratio test proposed by \cite{Be01}, denoted by $LR_{b,HARX-GJR}$, takes high values for all sixteen assets, suggesting that the null hypothesis is rejected with low p-values.

The results from the \cite{AmGi07} test are given in  Table \ref{AmGiTest}, which shows that our model provides better results overall, since the test statistic (\ref{AGTEST}) is almost always positive (the null hypothesis is not rejected at the 5\% level, mainly when we focus on the center or left side of the volatility distribution). In contrast, in the case of $AG_{RT}$, which assigns greater weight to the right tail of the distribution, the null hypothesis is rejected for ten assets.

\begin{table}[htb!]
\begin{center}
\centering
\small
\captionof{table}{Amisano-Giacomini test for the single asset analysis.} 
\begin{tabular}{cccccc}
\hline
Asset & $AG_{NW}$ & $AG_{CE}$ & $AG_{TL}$ & $AG_{RT}$ & $AG_{LT}$ \\
\hline
ATT & 1.3622 (0.1731) & 0.8815 (0.3780) & 1.6437 (0.1002) & 1.6545 (0.0980) & 0.8331 (0.4047) \\
BAC & 2.8124 (0.0049) & 2.1129 (0.0346) & 2.2914 (0.0219) & 3.1921 (0.0014) & 1.9912 (0.0464) \\ 
BOI & 0.3886 (0.6975) & -0.3921 (0.6943) & 1.2348 (0.2169) & 1.6890 (0.0913) & -0.7003 (0.4837) \\
CAT & 3.5401 (0.0003) & 3.0334 (0.0024) & 2.4433 (0.0145) & 3.4709 (0.0005) & 2.7339 (0.0062) \\
CTG & 2.5478 (0.0108) & 2.1816 (0.0291) & 2.1211 (0.0339) & 2.8752 (0.0040) & 1.7528 (0.0796) \\
FDX & 2.0846 (0.0371) & 1.4028 (0.1606) & 2.0505 (0.0403) & 2.6506 (0.0080) & 0.7403 (0.4591) \\
HON & 2.0001 (0.0454) & 1.1536 (0.2486) & 2.1087 (0.0349) & 2.6005 (0.0093) & 0.5590 (0.5761) \\
HPQ & 2.4106 (0.0159) & 1.9314 (0.0534) & 1.7141 (0.0865) & 2.7022 (0.0068) & 1.6046 (0.1085) \\
IBM & -0.6532 (0.5132) & -0.8718 (0.3833) & -0.4710 (0.6376) & 0.5245 (0.5999) & -0.7477 (0.4546) \\
JPM & 0.8703 (0.3841) & -0.0659 (0.9474) & 2.2998 (0.0214) & 1.7383 (0.0821) & -0.3541 (0.7232) \\
MDZ & 0.0482 (0.9615) & -0.5614 (0.5745) & 0.6841 (0.4939) & 1.1501 (0.2501) & -0.7592 (0.4477) \\
PEP & 2.3365 (0.0194) & 2.2043 (0.0275) & 1.4551 (0.1456) & 2.3862 (0.0170) & 1.6597 (0.0969) \\
PRG & 1.5118 (0.1305) & 1.0912 (0.2751) & 1.0372 (0.2996) & 1.5269 (0.1267) & 1.0288 (0.3035) \\
TWX & 1.1304 (0.2583) & 0.1237 (0.9015) & 2.1904 (0.0284) & 2.1361 (0.0326) & -0.4314 (0.6661) \\
TXN & 1.7525 (0.0796) & 1.0795 (0.2803) & 2.1059 (0.0352) & 2.3662 (0.0179) & 0.4049 (0.6855) \\
WFC & 1.4184 (0.1560) & 0.3584 (0.7200) & 2.5002 (0.0124) & 2.2908 (0.0219) & 0.0628 (0.9499) \\
\hline
\end{tabular}\par
\label{AmGiTest} 
\end{center}
\footnotesize{The table reports, for each stock (the ticker is given in the first column), the values of the likelihood ratio test (the p-values are given in brackets) proposed by \cite{AmGi07}, for different weights. Each weight places greater emphasis on particular regions of the distribution: center ($AG_{CE}$), tails ($AM_{TL}$), right tail ($AG_{RT}$) and left tail ($AG_{LT}$). $AG_{NW}$ coincides with the unweighted likelihood ratio test.}
\end{table}

Finally, the results of the \cite{DiMa02} test are given in Table \ref{DiMatest}, which shows that the sign of the test statistic is always negative, suggesting that our approach results in a lower loss $L_{\tau,t}\left(fpc_t,Q_{fpc_t}(\tau,\textbf{x}_{t-1})\right)$ overall. The performances are almost always statistically different, given that the null hypothesis is rejected at the 5\% level in nine of sixteen cases when we consider $\tau=0.1$ and in fifteen of sixteen cases with $DM_{0.5}$. When we consider $DM_{0.9}$, the null hypothesis is always rejected, with low p-values. Therefore, the three tests provide clear evidence that our model performs better, mainly in forecasting high levels of volatility.

\begin{table}[htb!]
\begin{center}
\centering
\small
\captionof{table}{Diebold-Mariano test for the single asset analysis.} 
\begin{tabular}{cccc}
\hline
Asset & $DM_{0.1}$ & $DM_{0.5}$ & $DM_{0.9}$ \\
\hline
ATT & -1.8480 (0.0000) & -4.6047 (0.0000) & -4.9845 (0.0000) \\
BAC & -1.3217 (0.1862) & -1.7453 (0.0809) & -1.9765 (0.0480) \\
BOI & -2.1425 (0.0321) & -5.2924 (0.0000) & -5.6195 (0.0000)\\
CAT & -1.4516 (0.1460) & -4.0947 (0.0000) & -3.9180 (0.0000) \\
CTG & -1.9613 (0.0498) & -3.0134 (0.0025) & -3.2932 (0.0009) \\
FDX & -2.0990 (0.0358) & -4.0106 (0.0000) & -4.9344 (0.0000) \\
HON & -1.7538 (0.0794) & -4.0245 (0.0000) & -4.3696 (0.0000) \\
HPQ & -1.7299 (0.0836) & -3.3081 (0.0009) & -3.2827 (0.0010) \\
IBM & -3.2205 (0.0012) & -4.9898 (0.0000) & -5.5552 (0.0000) \\
JPM & -1.9871 (0.0469) & -3.8176 (0.0001) & -3.7012 (0.0002) \\
MDZ & -2.7791 (0.0054) & -4.3246 (0.0000) & -5.9020 (0.0000) \\
PEP & -3.0889 (0.0020) & -5.2845 (0.0000) & -6.2248 (0.0000) \\
PRG & -3.4826 (0.0005) & -5.8129 (0.0000) & -6.2650 (0.0000) \\
TWX & -1.6146 (0.1063) & -3.2456 (0.0011) & -3.9691 (0.0001) \\
TXN & -0.9730 (0.3305) & -3.6680 (0.0002) & -3.1525 (0.0016) \\
WFC & -2.4493 (0.0143) & -4.1689 (0.0000) & -4.9362 (0.00007) \\
\hline
\end{tabular}\par
\label{DiMatest} 
\end{center}
\footnotesize{The table reports, for each stock (the ticker is given in the first column), the values of the likelihood ratio test (the p-values are given in brackets) proposed by \cite{DiMa02}, at $\tau=\{0.1,0.5,0.9\}$.}
\end{table}

To conclude, we found similar results between models (\ref{mainmodel}) and (\ref{single_asset_model}). In particular, for both models the lagged value of the response variable and the lagged value of the S\&P 500 return were fundamental explanatory variables at high $\tau$ levels, which are the most critical. In contrast, the lagged value of the jump component is significant in a few cases. We determined that the relationships between four explanatory variables ($rrv_{i,t-1}$, $vix_{t-1}$, $sp500_{t-1}$, and $jump_{i,t-1}$) and the conditional volatility quantiles are almost always stronger in model (\ref{mainmodel}) than in model (\ref{single_asset_model}). However, in the case of $\overline{rrv_5}_{i,t-1}$, the relationships are stronger in model (\ref{single_asset_model}) than in model(\ref{mainmodel}). Finally, even in the single-asset analysis, the goodness of the predicted power of our approach is validated by means of the three tests.

\section{Conclusions}

We proposed a method by which to model and forecast the 
conditional distribution of asset returns’ volatility. We used the quantile regression approach, considering as predictors variables built from the lagged values of the estimated volatility, following the HAR structure \cite{Co09} developed, and macroeconomic and financial variables that reflect the overall market behavior.  

We estimated volatility using the realized range-based bias corrected bipower variation introduced by \cite{ChPoVe09}, which is a consistent estimator of the integrated variance in the presence of microstructure noise and jumps in the context of high-frequency data. 
Our analyses considered sixteen companies that operate in a variety of sectors in the U.S. market, and the results provide evidence of relevant impacts by the explanatory variables. In particular, the lagged values of the estimated volatility and the  S\&P 500 return were critical indicators in the context of extreme events, where volatility can reach considerably high levels. These two regressors were highly significant in terms of their ability to explain the high quantiles of volatility. Moreover, the test \cite{KoBa82} introduced allowed us to reject the location-shift hypothesis, highlighting the heterogeneous impacts of the regressors across quantiles. 

In order to assess the evolution of the relationships among the variables over time, we carried out a rolling analysis with steps of one day and subsamples consisting of 500 observations. Thus, verified that two special events, the U.S. subprime crisis and the European sovereign debt crisis, have affected those relationships. In particular, acute sensitivity was recorded at high levels of quantiles. Finally, the tests developed by \cite{Be01}, \cite{AmGi07} and  \cite{DiMa02} validated the forecast performances, even in periods of financial turmoil. We compared our approach with a HARX-GJR model, which combines a HAR structure on the realized volatility mean with additional exogenous variables, and a GJR-GARCH \citep{GlJaRu93} for the mean innovation variances. The results confirm the superior performance of our approach mainly when we focus on the right tail of the volatility distribution, which assumes critical importance in our framework.

Our findings provide supporting evidence for the use of quantile regression methods for the quantile forecasts and for the density forecast of the realized range volatility. The improvement over traditional methods is marked and will be relevant in all areas where volatility quantile values and volatility density forecast play a role.

\vspace{1cm}
\textbf{Acknowledgments}\\
We wish to thank Fabio Busetti, Sandra Paterlini 
and the participants to the ``2nd CIdE Workshop for PhD students in Econometrics and Empirical Economics'', organized by Centro Interuniversitario di Econometria, and to the seminar at the EBS University of Wiesbaden (Germany), organized by the Chair of Financial Econometrics and Asset Allocation. Massimiliano Caporin acknowledges financial support from the MIUR PRIN project MISURA - Multivariate Statistical Models for Risk Assessment. Both
authors acknowledge financial support from the University of Padova. Usual disclaimers apply.

\bibliographystyle{Chicago}
\bibliography{mybib}

\clearpage

\appendix

\section{Volatility estimation in the presence of noise and jumps}

Let $p_t$ be the logarithmic price of a financial asset at time $t$. We assume that it follows the Brownian semi-martingale process:

\begin{equation} \label{log_pri}
p_t=p_0+\int_{0}^{t} \mu_u d u +\int_{0}^{t} \sigma_u d W_u,\:\;\;\;t \geq 0,
\end{equation} 

where the drift $\mu=(\mu_t)_{t\geq 0}$ is locally bounded and predictable, and $\sigma=(\sigma_t)_{t \geq 0}$ is a strictly positive process, independent of the standard Brownian Motion $W=(W_t)_{t \geq 0}$, and c\`adlag.  

In the high-frequency context, the quadratic variation assumes an important role. If a trading day equals the interval $[0,1]$ and is divided into \textit{n} subintervals with same width, that is, $0=t_0<t_1<...<t_n=1$, the  quadratic variation is defined as
\begin{equation}
QV=\plim_{n \rightarrow \infty} \sum_{i=1}^{n}(p_{t_i}-p_{t_{i-1}})^2,
\end{equation}

with $\max_{1 \leq i \leq n}\{t_i-t_{i-1}\}\rightarrow 0$. If the price evolution is described by equation (\ref{log_pri}), and $\mu_u$ and $\sigma_u$ satisfy certain regularity conditions, the quadratic variation equals the integrated volatility \citep{HuWh87}:

\begin{equation}
IV=\int_{0}^{t}\sigma^2_u d u.
\end{equation}
 
Let $m$ be the number of prices recorded at each subinterval and $N$ the total number of observations for a trading day, that is, $N=mn$. The daily volatility can be estimated through the realized variance:

\begin{equation} \label{realvar}
RV^N=\sum_{i=1}^{N}r^2_{i\Delta,\Delta},
\end{equation}

where $r_{i\Delta,\Delta}=p_{i/N}-p_{(i-1)/N}$ is the intraday return recorded at the $i$-th discrete point for $i=1,...,N$, and $\Delta=1/N$. If microstructure noise is absent, $RV^N$ is a consistent estimator of $IV$ as $N \rightarrow \infty$. In particular, \cite{Ja94}, \cite{JaPr98}, and \cite{BaNiSh02} obtained the asymptotic distribution of $RV^N$: 

\begin{equation}
N^{1/2}\left(RV^N-\int_{0}^{1}\sigma_u^2 du \right) \stackrel{d}{\rightarrow} \mathcal{MN} \left(0,2\int_{0}^{1} \sigma_u^4 d u\right),
\label{den1}
\end{equation}

where $\mathcal{MN}$ denotes the mixed normal distribution. In (\ref{den1}) $\int_{0}^{1} \sigma_u^4 d u$ is the integrated quarticity ($IQ$), which can be estimated through the realized quarticity, denoted as $RQ^N$. The realized quarticity is given as

\begin{equation}
RQ^N=\frac{N}{3}\sum_{i=1}^{N}r^4_{i \Delta,\Delta} \stackrel{p}{\rightarrow} \int_{0}^{1}\sigma_u^4 d u.
\end{equation}

We stress that $RV^N$ is computed by considering just the last price of each subinterval. In order to reduce this information loss (within interval prices are completely disregarded) \cite{MaVD07} and \cite{ChPo07} proposed the realized range-based variance ($RRV^{n,m}$), a modified version of the quantity given in equation (\ref{realvar}). Thus, more information is used, as the maximum and the minimum prices are both taken into account in each subinterval. Let $s_{p_{i\Delta,\Delta},m}=\max_{0 \leq s,t \leq m}\left( p_{\frac{i-1}{n}+\frac{t}{N}}-p_{\frac{i-1}{n}+\frac{s}{N}} \right)$ be the range for $i=1,...,n$; the estimator of interest is defined as:

\begin{equation} \label{real_range}
RRV^{n,m}=\frac{1}{\lambda_{2,m}}\sum_{i=1}^{n}s^2_{p_{i\Delta,\Delta},m},
\end{equation}

where  $\lambda_{r,m}=E\left[ \left| \max_{0\leq s,t \leq m}\left(W_{\frac{t}{m}}-W_{\frac{s}{m}}\right)\right|^r\right]$ is the $r$th moment of the range of a standard Brownian Motion ($W$) over a unit interval; $\lambda_{r,m}$ is computed through numerical simulation and $\lambda_{2,m} \to \lambda_2=4 \log (2)$ as $m \to \infty$. \cite{ChPoVe09} showed that, without microstructure noise, $RRV^{n,m}\stackrel{p}{\rightarrow} \int_{0}^{1}\sigma_u^2 du$ as $n\rightarrow \infty$ and 

\begin{equation}
\sqrt{n}\left(RRV^{n,m}-\int_{0}^{1}\sigma^2_u d u\right) \stackrel{d}{\rightarrow} \mathcal{MN} \left(0,\Lambda_c \int_{0}^{1}\sigma_u^4 d u \right),
\label{den2}
\end{equation}

where $\Lambda_c=\lim_{m \to c} \Lambda_m$ and $\Lambda_m=(\lambda_{4,m}-\lambda_{2,m}^2)/\lambda_{2,m}^2$; $\Lambda_m$ is decreasing in $m$ and takes values between 2 ($m=1$) and about 0.4 ($m\rightarrow \infty$). Therefore, comparing (\ref{den1}) and (\ref{den2}) shows that $RRV^{n,m}$ is more efficient than $RV^N$ if $m>1$. 

So far, we have not considered microstructure noise and the effects of price jumps, although they might have a significant impact. As a consequence, the estimators described above might be biased because of the effects of measurement errors. Suppose that $p_t$ satisfies equation (\ref{log_pri}) and that the new price process is equal to $p^*_t=p_t+\eta_t$, with $\eta$ denoting the microstructure noise. \cite{ChPoVe09} modeled the noise $\eta=\left( \eta_t\right)_{t \geq 0}$ as a sequence of i.i.d. random variables, such that $\mathbb{E}[\eta_t]=0$, $\mathbb{E}[\eta_t^2]=\omega^2$, and $\eta \perp p$, showing that 

\begin{equation}
\hat{\omega}^2_N=\frac{RV^N}{2N}\stackrel{p}{\rightarrow} \omega^2,
\end{equation} 

and

\begin{equation}
N^{1/2}(\hat{\omega}^2_N-\omega^2)\stackrel{d}{\rightarrow} \mathcal{N}(0,\omega^4).
\end{equation}

Moreover, it is assumed that the jump component affects the price process through the relationship $\tilde{p}_t=p^*_t+\sum_{i=1}^{N_t}J_i$, where $J=(J_i)_{i=1,...,N_t}$ is the jump size component and $N=(N_t)_{t \geq 0}$ is a finite activity-counting process. In particular, 
in order to test the null hypothesis of no jumps at day $t$, \cite{RePEc:zbw:sfb475:200637} proposed the following statistic:
\begin{equation}\label{test_ZT,t}
Z_{TP,t}=\frac{\sqrt{n}\left(1-RBV^{n,m}_t/RRV^{n,m}_t\right)}{\sqrt{\nu_m \max\left\lbrace RQQ^{n,m}_t / \left( RBV^{n,m}_t\right)^2,1\right\rbrace}} \stackrel{d}{\rightarrow} \mathcal{N}(0,1).
\end{equation}
In equation (\ref{test_ZT,t}), $RBV^{n,m}$ denotes the range-based bipower variation, whereas 
$RQQ^{n,m}$ is the range-based quad-power quarticity, defined, respectively, as:
\begin{equation}
RBV^{n,m}=\frac{1}{\lambda_{2,m}}\sum_{i=1}^{n-1}s_{p_{i\Delta,\Delta,m}}s_{p_{(i+1)\Delta,\Delta,m}},
\end{equation}
and
\begin{equation}
RQQ^{n,m}=\frac{n}{\lambda^4_{1,m}}\sum_{i=1}^{n-3}s_{p_{i\Delta,\Delta,m}}s_{p_{(i+1)\Delta,\Delta,m}}s_{p_{(i+2)\Delta,\Delta,m}}s_{p_{(i+3)\Delta,\Delta,m}},
\end{equation}
where $\nu_m=\lambda^2_{2,m}\left(\Lambda^R_m + \Lambda^B_m - 2 \Lambda^{RB}_m\right)$, $\Lambda^R_m= \left( \lambda_{4,m}-\lambda^2_{2,m}\right)/\lambda^2_{2,m}$, $\Lambda^B_m=( \lambda^2_{2,m}+2\lambda^2_{1,m}\lambda_{2,m}-3 \lambda^4_{1,m}) /\lambda^4_{1,m}$, $\Lambda^{RB}_m= (2 \lambda_{3,m} \lambda_{1,m}- 2 \lambda_{2,m} \lambda^2_{1,m})/ (\lambda_{2,m}\lambda^2_{1,m})$. 

Finally, \cite{ChPoVe09} proposed the realized range-based bias corrected bipower variation ($RRV_{BVBC}^{n,m}$), a consistent estimator of the integrated variance in the presence of price jumps and noise; the estimator is given in the following equation:  

\begin{equation}\label{RRVBVBC}
RRV_{BVBC}^{n,m}=\frac{1}{\tilde{\lambda}_{1,m}^2}\sum_{i=1}^{n-1}\left| s_{p_{i\Delta,\Delta},m}-2\hat{\omega}_N \right| \left| s_{p_{(i+1)\Delta,\Delta},m}-2\hat{\omega}_N \right|,
\end{equation}

where $\tilde{\lambda}_{r,m}=\mathbb{E}\left[\left| \max_{t:\eta \frac{t}{m}=\omega,\;s:\eta\frac{s}{m}=-\omega}\left(W_{\frac{t}{m}}-W_{\frac{s}{m}}\right)\right|^r\right]$, with $1 \leq s,t\leq m$. $RRV_{BVBC}^{n,m}$ assumes an important role in the present work since it is used to estimate the volatility in the model presented in Section 3.

\section{Quantile regression: estimation and testing}

The method of least squares is a widely used tool in statistics, given its attractive computational tractability, its simplicity, and the optimal results it guarantees given certain assumptions. Nevertheless, this approach might lead to erroneous conclusions when the model’s hypotheses are violated. Here we concentrate on one relevant assumption, that of linearity.

In this case, \cite{KoBa78} proposed an alternative method, regression quantiles, that ensures robust results. Regression quantiles allows one quantile of the conditional distribution of a variable to be estimated instead of restricting attention to the conditional mean. Let $Y$ be a real-valued random variable with distribution function $F_Y(y)=P(Y \leq y)$. For any $0<\tau <1$, the $\tau$-th quantile of $Y$ is equal to $F_Y^{-1}(\tau)=\inf\{y:F_Y(y)\geq \tau \}$. 

The approach introduced by \cite{KoBa78} makes use of the asymmetric loss function,

\begin{equation}
\rho_\tau(u)=u[\tau-I(u<0)],
\end{equation}

showing that the minimizer $\tilde{y}_{\tau}$ of the expected loss function $\mathbb{E}[\rho_\tau(Y-\tilde{y}_{\tau})]$ satisfies $F_Y(\tilde{y}_{\tau})-\tau=0$. In particular, $\tilde{y}_{\tau}$ is the conditional quantile function $Q_Y(\tau|X_1,X_2,...,X_k)$ in the linear quantile regression: 

\begin{equation} \label{ytsz}
\tilde{y}_{\tau}=Q_Y(\tau|X_1,X_2,...,X_k)=\beta_0(\tau)+\beta_1(\tau)X_1+...+\beta_k(\tau)X_k,
\end{equation}

where $\textbf{X}=(X_1,X_2,...,X_k)$ is a vector with $k$ explanatory variables. 

Given the time index $t=1,...,T$, let $y_t$ and $x_{i,t}$ be the realization of the random variable $Y$ and the observed value of the $i$-th explanatory variable, respectively, for $i=1,...,k$, at day $t$. Then the parameter vector $\hat{\pmb{\beta}}(\tau)=(\hat{\beta_0}(\tau),...,\hat{\beta}_k(\tau))$ in (\ref{ytsz}) is estimated as a solution of the quantile regression problem:

\begin{equation}\label{distfunct}
 \operatorname*{min}_{\mathbf{\beta} \in \mathbb{R}^{k+1}}\sum_{t=1}^{T}\rho_\tau(y_t-\beta_0-\beta_1x_{1,t}-...-\beta_kx_{k,t}).
\end{equation} 

Let $\mathbf{x}_t$ be the row vector that includes the $t$-th observation of $\mathbf{X}$ (if the errors $\epsilon_t=y_t-\textbf{x}'_t \pmb{\beta}(\tau)$ are i.i.d.), with distribution $F_{\epsilon}$ and density $f_{\epsilon}$, where $f_{\epsilon}\left(F_{\epsilon}^{-1}(\tau)\right)>0$ is in the neighborhood of $\tau$, and the quantile estimator $\hat{\pmb{\beta}}(\tau)$ is asymptotically distributed as

\begin{equation} \label{asymp_1}
\sqrt{T}\left[\hat{\pmb{\beta}}(\tau)-\pmb{\beta}(\tau)\right]\stackrel{d}{\rightarrow} \mathcal{N}\left( 0,\tilde{k}^2(\tau)\mathbf{D}^{-1}\right),
\end{equation}

where $\tilde{k}^2(\tau)=\frac{\tau(1-\tau)}{f_{\epsilon}\left(F_{\epsilon}^{-1}(\tau)\right)^2}$ and $\mathbf{D}=\lim_{T \to \infty}\frac{1}{T}\sum_{t}\mathbf{x}_t'\mathbf{x}_t$ is a positive definite matrix. 

When the errors are independent but non-identically distributed, the error density $f_{\epsilon,t}$ changes in the sample. In this context, the asymptotic distribution becomes

\begin{equation} \label{asymp_2}
\sqrt{T}\left[\hat{\pmb{\beta}}(\tau)-\pmb{\beta}(\tau)\right]\stackrel{d}{\rightarrow} \mathcal{N}\left(0,\tau(1-\tau)\mathbf{D}_1(\tau)^{-1}\mathbf{D}\mathbf{D}_1(\tau)^{-1}\right),
\end{equation}

where $\mathbf{D}_1(\tau)=\lim_{T \to \infty}\frac{1}{T}\sum_{t}f_{\epsilon_,t}(F_{\epsilon}^{-1}(\tau))\mathbf{x}_t'\mathbf{x}_t$ is a positive definite matrix \citep{KoBa82}.

Equations (\ref{asymp_1}) and (\ref{asymp_2}) show that the asymptotic covariance matrix of the quantile regression coefficients depends on the error density, so the matrix could be difficult to estimate. This problem can be avoided by developing the inferential procedure in a different way. Resampling methods to estimate the parameters' standard errors are a valid alternative to the asymptotic results discussed above, and several studies recommend using the bootstrap method in the quantile regression framework (e.g., \cite{Bu95}). \cite{Ef79} introduced the computer-based bootstrap method to estimate the variance and distribution of an estimate and, more generally, of a statistic. The main advantages of the bootstrapping approach are well-known: it assumes no particular distribution of the errors, it is not based on asymptotic model properties, and it is available regardless of the statistic of interest’s complexity. \cite{Ef79} showed that this approach works well on a variety of estimation problems. Nevertheless, the bootstrap estimates are influenced by the sample’s variability and the bootstrap resampling’s variability, the former from the fact that the estimates are based on just one sample for a certain population and the latter from the finite number of replications \citep{DaFuVi14}. 

Several types of bootstrapping methods are discussed in the literature and applied in quantile regression estimation. Three of them are the $xy$-pair method \citep{Koc03}, the method based on pivotal estimating functions \citep{PaWeYi94}, and the Markov chain marginal bootstrap \citep{HeHu02}. \cite{DaFuVi14} compared these three techniques by using two kinds of models: a quantile regression homogeneous error model, in which the error term does not depend on the explanatory variables, and a quantile regression heterogeneous error model, in which the error term is a function of the regressors. Using simulated data, the authors show that the three methods produce similar results for a homogeneous model in terms of estimated coefficients and standard errors. However, when the heterogeneous model is considered, the $xy$-pair method has the best results, while the worst results come from the Markov chain marginal bootstrap. We use the $xy$-pair method to estimate the parameters' standard errors in our work.

When time series are used, the problem of serial correlation becomes critical, a relevant issue in the present work, whose aim is to model the conditional quantiles of volatility, a variable whose current value is typically affected by past values. If the dynamic of the model is neglected, the errors become serially correlated, with consequences in inference. Several studies have applied the quantile regression, taking into account models with autoregressive structure. For instance, \cite{KoXi04} proposed the Quantile Autoregression (\textit{QAR}) model, in which the $\tau$-th conditional quantile of the response variable is explained by the lagged values of the same dependent variable. The authors focused on the first-order autoregression (although the analysis could be extended to the general case) and estimated the parameters of the model by solving the problem given in equation (\ref{distfunct}), using as regressor the lagged value of the response variable. \cite{EnMa04} proposed the CAViaR model to estimate the conditional Value-at-Risk of a financial institution. The model has an autoregressive structure and is estimated following the approach proposed by \cite{KoBa78}. \cite{We90} performed a median regression for a model with serial correlation and found that the estimates are unbiased but the inference is wrong. Therefore, it is important to check for residual serial correlation and to take into account some possible solutions if such correlations are present, such as adding the lagged values of the variables or computing the parameters' standard errors by means of the bootstrapping methods mentioned above.

A further aspect of the evaluation of the quantile regression output refers to the goodness-of-fit assessment. \cite{KoMa99} introduced a goodness-of-fit quantity for quantile regression that is analogous to the coefficient of determination for the least squares regression. For simplicity, we introduce the approach in a quantile regression model for $y_t$ with just one regressor, $x_t$. At a given quantile $\tau$, we evaluated two quantities, the residual absolute sum of weighted differences, denoted as

\begin{eqnarray}
RASW_\tau & = & \sum_{y_t \geq \hat{\beta}_0(\tau)+\hat{\beta}_1(\tau)x_t}\tau \left| y_t-\hat{\beta}_0(\tau)-\hat{\beta}_1(\tau)x_t\right|  \nonumber \\ & + & \sum_{y_t < \hat{\beta}_0(\tau)+\hat{\beta}_1(\tau)x_t}(1-\tau) \left| y_t-\hat{\beta}_0(\tau)-\hat{\beta}_1(\tau)x_t\right|,
\end{eqnarray}

and the total absolute sum of the weighted differences, which reads

\begin{eqnarray}
TASW_\tau=\sum_{y_t \geq \hat{q}_y(\tau)}\tau \left| y_t - \hat{q}_y(\tau) \right|+\sum_{y_t < \hat{q}_y(\tau)}(1-\tau) \left| y_t - \hat{q}_y(\tau) \right|,
\end{eqnarray}

where $\hat{q}_y(\tau)$ is the estimated unconditional $\tau$th quantile of $Y$. Given these two quantities, a pseudo $R^2$ is defined as:

\begin{equation}\label{pseudo_R2}
R^1(\tau)=1-\frac{RASW_\tau}{TASW_\tau}.
\end{equation}

Notably, $R^1(\tau)$ ranges between 0 and 1, like the coefficient of determination. Nevertheless, it is a local measure of fit, so, unlike the $R^2$ in the least squares regression context, it can't be used as a global goodness-of-fit measure because it quantifies the relative success of two models, restricted and unrestricted, at a given quantile \citep{KoMa99}. 

Another approach to evaluating the goodness of fit at a specified quantile is that proposed by \cite{KoBa82b}. In that study’s testing framework, the null hypothesis is that a set of explanatory variables used to specify a conditional quantile in a general model does not improve the fit with respect to a restricted model (where those variables are not included). Therefore, the test reads like the F-test for the significance of a subset of coefficients. In the quantile regression framework, the test is of a Wald-type, and the associated test statistic, $\xi_w$, is based on the estimated coefficients of the unrestricted model.  

In a quantile regression approach, several hypotheses can be tested on the conditional quantile parameters or on the innovations. One of these hypotheses is of particular interest for the analysis of realized variance series. We refer to the so-called \textit{location shift hypothesis}, which requires the parameters that multiply the explanatory variables in (\ref{ytsz}) to be identical over $\tau$. Thus, changes among the conditional quantiles occur only in the intercepts. If this null hypothesis is rejected, (\ref{ytsz}) would be a location-shift and scale-shift model. In the present work, the location-shift hypothesis is tested by means of a variant of the Wald test that was introduced by \cite{KoBa82}. The null hypothesis of the test is that the coefficient slopes are the same across quantiles. The test statistic is asymptotically distributed as $\mathcal{F}$, and the numerator’s degrees of freedom are equal to the rank of the null hypothesis, while those related to the denominator are determined by subtracting the sample size by the number of parameters that characterize the model of interest. Clearly, if the location-shift hypothesis is accepted, the potential advantages of quantile regressions over linear regressions would be limited, as the covariates would have constant impact across the response quantiles.

One of the most important properties of the approach introduced by \cite{KoBa78} is the robustness of the results to some forms of non-linearity such that it overcomes the gaps related to the least squares framework. 

The quantile regression approach allows several conditional quantiles to be estimated, but the estimation focuses on a single quantile. Therefore, once the estimates of many quantiles are separately performed, whether conditional quantiles are coherent (or correctly specified) and do not overlap must be verified. For instance, we must determine that the predicted 95th percentile of the response is higher than the 90th percentile. This check guarantees the appropriateness of the conditional distribution that can be recovered from the estimation of a large number of conditional quantiles. If quantiles cross, corrections may be in order. For instance, \cite{Ko84} applied parallel quantile planes to rectify this problem.


\end{document}